\documentclass[twocolumn,prl,10pt,aps,floatfix]{revtex4-1}
\usepackage[english]{babel}
\usepackage[colorlinks,linkcolor=blue,citecolor=blue,urlcolor=blue]{hyperref}
\usepackage{amsmath}
\usepackage{amssymb}
\usepackage{graphicx}
\begin{document}
\title{Block-Spiral Magnetism: An Exotic Type of Frustrated Order}
\author{J. Herbrych$^{1,2,3}$}
\author{J. Heverhagen$^{4,5}$}
\author{G. Alvarez$^{6}$}
\author{M. Daghofer$^{4,5}$}
\author{A. Moreo$^{1,2}$}
\author{E. Dagotto$^{1,2}$}
\affiliation{$^{1}$ Department of Physics and Astronomy, University of Tennessee, Knoxville, Tennessee 37996, USA}
\affiliation{$^{2}$ Materials Science and Technology Division, Oak Ridge National Laboratory, Oak Ridge, Tennessee 37831, USA}
\affiliation{$^{3}$ Department of Theoretical Physics, Faculty of Fundamental Problems of Technology, Wroc{\l}aw University of Science and Technology, 50-370 Wroc{\l}aw, Poland}
\affiliation{$^{4}$ Institute for Functional Matter and Quantum Technologies, University of Stuttgart, Pfaffenwaldring 57, D-70550 Stuttgart, Germany}
\affiliation{$^{5}$ Center for Integrated Quantum Science and Technology, University of Stuttgart, Pfaffenwaldring 57, D-70550 Stuttgart, Germany}
\affiliation{$^{6}$ Computational Sciences and Engineering Division and Center for Nanophase Materials Sciences, Oak Ridge National Laboratory, Oak Ridge, Tennessee 37831, USA}
\date{\today}
%================================================================================
\begin{abstract}
Competing interactions in Quantum Materials induce novel states of matter such as frustrated magnets, an extensive field of research both from the theoretical and experimental perspectives. Here, we show that competing energy scales present in the low-dimensional orbital-selective Mott phase (OSMP) induce an exotic magnetic order, never reported before. Earlier neutron scattering experiments on iron-based 123 ladder materials, where OSMP is relevant, already confirmed our previous theoretical prediction of block-magnetism (magnetic order of the form $\uparrow\uparrow\downarrow\downarrow$). Now we argue that another novel phase can be stabilized in multi-orbital Hubbard models, the {\it block-spiral state}. In this state, the magnetic islands form a spiral propagating through the chain but with the blocks maintaining their identity, namely rigidly rotating. This new spiral state is stabilized without any apparent frustration, the common avenue to generate spiral arrangements in multiferroics. By examining the behaviour of the electronic degrees of freedom, parity breaking quasiparticles are revealed. Finally, a simple phenomenological model that accurately captures the macroscopic spin spiral arrangement is also introduced, and fingerprints for the neutron scattering experimental detection of our new state are provided.
\end{abstract}
%================================================================================
\maketitle
%================================================================================

Frustrated magnetism is one of the main areas of research in contemporary Condensed Matter Physics. In the generic scenario, magnetic frustration emerges from the failure of the system to fulfill simultaneously conflicting local requirements. As a consequence, complex spin patterns can develop from geometrical frustration (as in triangular, Kagome, or pyrochlore lattices) or from special spin-spin interactions (long-range exchange, Dzyaloshinskii–Moriya coupling, Kitaev model, spin-orbit effects, and others). In real materials both scenarios often coexist. Competing mechanisms can lead to interesting phenomena, such as spiral order \cite{McCulloch2008,Sato2011,Soos2016}, spin ice \cite{Gingras2014}, skyrmions \cite{Nagaosa2013}, spin liquids \cite{Tsunetsugu1992,Balents2010}, and resonating valence bond states \cite{Kimchi2018}. Also, the electronic properties of such systems are of much interest: it was shown that the interplay of a spiral state on a metallic host can support Majorana fermions \cite{Oreg2010,Braunecker2013,Klinovaja2013,Perge2014,Schecter2015,Steinbrecher2018} and can also induce multiferroicity \cite{Sergienko2006,Dong2008,Seki2008,Furukawa2010,Tokura2010,Povarov2016,Scaramucci2018}. 

Another example of competing interactions is the orbital-selective Mott phase (OSMP) \cite{Medici2009,Caron2012}. In multi-orbital systems, this effect can lead to the selective localization of electrons on some orbitals. The latter coexist with itinerant bands of mobile electrons. This unique mixture of localized and itinerant components in multi-orbital systems could be responsible for the (bad) metallic behaviour of the parent compounds of iron-based superconductors. This is in stark contrast to cuprates, usually described by the single-band Hubbard model, where parent materials are insulators. Furthermore, the OSMP can also host exotic magnetic phases. It was shown that the competition between Hund and Hubbard interaction can stabilize unexpected block-magnetism \cite{Garcia2004,Garcia2010,Rincon2014,Herbrych2018,Herbrych2019}, namely antiferromagnetically (AFM) coupled ferromagnetic (FM) islands. The size and shape of such islands depends on the electronic filling of the itinerant orbitals. This work focuses on two representative cases: (i) the \mbox{$\pi/2$-block} spin-pattern $\uparrow\uparrow\downarrow\downarrow$ and (ii) the \mbox{$\pi/3$-block} spin-pattern $\uparrow\uparrow\uparrow\downarrow\downarrow\downarrow$. Note that these block-patterns are not spin-density waves: the local expectation values of spin operators yield uniform magnetization throughout the system, unlike a spin wave that would have peaks and valleys. Our study, on the other hand, indicates a clear block structure with spins of the same magnitude at each site \cite{Rincon2014,Herbrych2018,Herbrych2019}. Moreover, exact diagonalization results on small lattices \cite{Herbrych2018} indicate that the block-OSMP ground state has a large overlap (at least $50\%$) with a state of the form \mbox{$|\uparrow\uparrow\downarrow\downarrow\rangle-|\downarrow\downarrow\uparrow\uparrow\rangle$} (as exemplified for $\pi/2$-block). As a consequence, our states can be viewed as N\'eel-like states of an {\it enlarged magnetic unit cell}.

The OSMP was shown \cite{Caron2012} to be relevant for the low-dimensional family of iron-based ladders, the so-called 123 family AFe$_2$X$_3$, where A=Ba, K, Rb is an alkaline earth metal and X=S, Se is a chalcogen. From the magnetism perspective, two phases were experimentally reported: (i) for BaFe$_2$Se$_3$ inelastic neutron scattering (INS) identified \cite{Mourigal2015} a 2$\times$2 block-magnetic phase in a $\uparrow\uparrow\downarrow\downarrow$-pattern along the legs. Neutron diffraction measurements and muon spin relaxation yield the same conclusion \cite{Wu2019}. (ii) On the other hand, for BaFe$_2$S$_3$ and RbFe$_2$Se$_3$ INS reported \cite{Takahashi2015,Wang2016,Wang2017} 2$\times$1 blocks, FM along rungs and AFM along legs. Both of these phases are captured by the multi-orbital Hubbard Hamiltonian \cite{Rincon2014} and also by its low-energy effective description \cite{Herbrych2019}, the generalized Kondo-Heisenberg model.

\begin{figure}[!htb]
\includegraphics[width=1.0\columnwidth,]{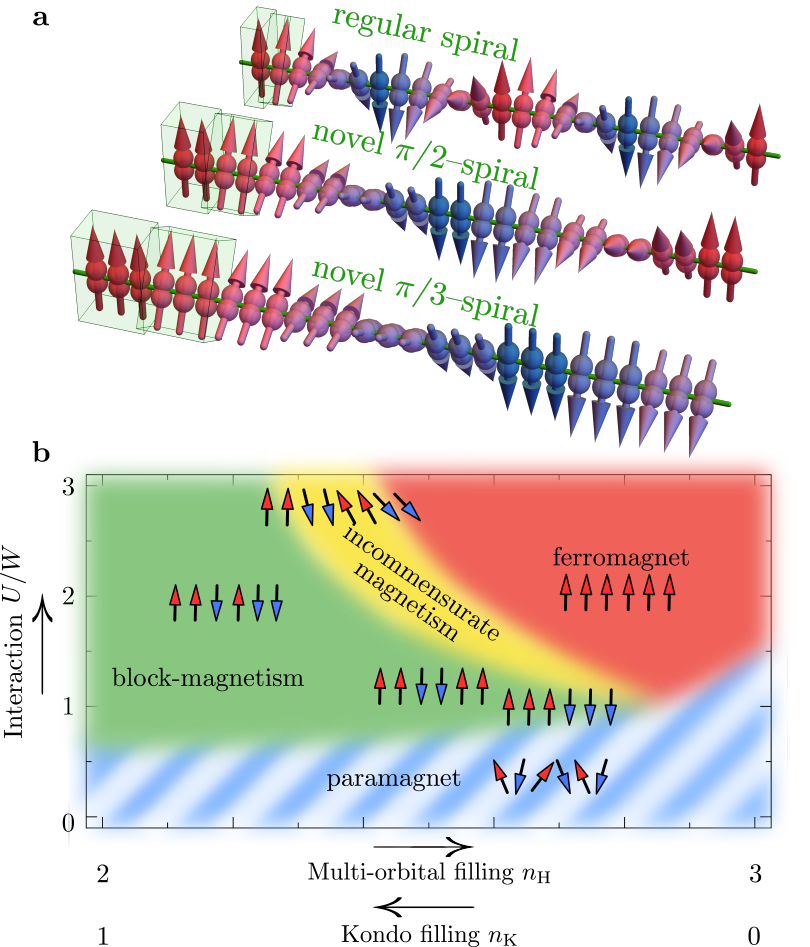}
\caption{Schematic representation of spirals within the orbital-selective Mott phase. ({\bf a}) Top to bottom: {\it standard spiral} spin structure with site-to-site spins rotation, novel {\it block-spirals} of two and three sites, respectively. ({\bf b}) Interaction $U$--filling $n_{\mathrm{H/K}}$ phase diagram. Solid (dashed) coloring represents the OSMP (paramagnetic) region.}
\label{schematic}
\end{figure}

In this work, we will unveil another novel magnetic phase that we stabilized, namely we will report an exotic {\it block-spiral} state. Such a state arises as a consequence of simultaneous tendencies in the system to form magnetic blocks and to develop non-collinear order. Different from {\it standard} spirals where one can observe spin-to-spin rotation, in the block-spiral state the blocks rigidly rotate, as illustrated in the three panels of Fig.~\ref{schematic}(a)]. The block spiral states we report have similarities with states in the rare-earth material TbFeO$_3$ \cite{Artyukhin2012} that display spin incommensurability and domain walls. However, the length scales in TbFeO$_3$ are much larger 340{\AA} and magnetic fields and anisotropy are needed for their stabilization. Furthermore, our new magnetic pattern appears without any apparent frustration in the model and is a consequence of subtly competing energy scales present in the OSMP. This novel spin order originates in ferromagnetic tendencies, induced by Hund physics in multi-orbital systems, and opposite antiferromagnetic superexchange tendencies, caused by Hubbard interactions. There is a {\it hidden frustration} in the system, not obvious at the Hamiltonian level, and whose exotic consequences appear only when powerful computational tools are used: simpler biased techniques likely would have missed this new state. We propose two simpler phenomenological models that capture the essence of our findings. Moreover, the electronic properties can be accurately described by quasiparticles that break parity symmetry, as expected within a spiral state. Also, we will show that the behaviour of spins can be effectively modeled via a frustrated long-range spin-only Heisenberg model. Such a spin model can serve as a starting point for the spin-wave theory calculations often used to compare theory vs INS spectra. Our findings are robust against system parameter changes and should characterize generic multi-orbital systems in the OSMP regime, close to ferromagnetism.

%--------------------------------------------------------------------------------
%--------------------------------------------------------------------------------

%--------------------------------------------------------------------------------
\vspace{1em}
\noindent {\bf \large Results}
\vspace{1em}
%--------------------------------------------------------------------------------

{\bf A. Magnetism of OSMP.} We will discuss the properties of a multi-orbital Hubbard model on a one-dimensional (1D) lattice. In the generic SU(2)--symmetric form it can be written as
\begin{eqnarray}
H_{\mathrm{H}}&=&-\sum_{\gamma,\gamma^\prime,\ell,\sigma}
t_{\gamma\gamma^\prime}
\left(c^{\dagger}_{\gamma,\ell,\sigma}c^{\phantom{\dagger}}_{\gamma^\prime,\ell+1,\sigma}+\mathrm{H.c.}\right)+
\Delta\sum_{\ell}n_{1,\ell}\nonumber\\
&+&U\sum_{\gamma,\ell}n_{\gamma,\ell,\uparrow}n_{\gamma,\ell,\downarrow}
+\left(U-5J_{\mathrm{H}}/2\right)\sum_{\ell}n_{0,\ell}n_{1,\ell}\nonumber\\
&-&2J_{\mathrm{H}}\sum_{\ell}\mathbf{S}_{0,\ell} \cdot \mathbf{S}_{1,\ell}
+J_{\mathrm{H}}\sum_{\ell}\left(P^{\dagger}_{0,\ell}P^{\phantom{\dagger}}_{1,\ell}
+\mathrm{H.c.}\right)\,.
\label{hamhub}
\end{eqnarray}
Here, $c^{\dagger}_{\gamma,\ell,\sigma}$ ($c^{\phantom{\dagger}}_{\gamma,\ell,\sigma}$) creates (destroys) an electron with spin projection $\sigma=\{\uparrow\,,\downarrow\}$ at orbital $\gamma=\{0,1\}$ of site $\ell=\{1\,,\dots,L\}$. $\Delta$ stands for crystal-field splitting, while $n_{\gamma,\ell}=\sum_{\sigma}n_{\gamma,\ell,\sigma}$ represents the total density of electrons at $(\gamma,\ell)$, with $n_{\gamma,\ell,\sigma}$ the density of electrons with $\sigma$-spin projection. $U$ is the standard, same-orbital repulsive Hubbard interaction, and $J_\mathrm{H}$ is the Hund exchange between spins $\mathbf{S}_{\gamma,\ell}$ at different orbitals $\gamma$. Finally, the last term $P^{\dagger}_{0,\ell}P^{\phantom{\dagger}}_{1,\ell}$ stands for pair-hopping between orbitals, where $P_{\gamma,\ell}=c^{\phantom{\dagger}}_{\gamma,\ell,\uparrow}c^{\phantom{\dagger}}_{\gamma,\ell,\downarrow}$.

In the most generic case, the Fe-based materials should be modeled with five $3d$-orbitals (three \mbox{$t_{2g}$}: $d_{xy}$, $d_{yz}$, $d_{xz}$, and two \mbox{$e_g$}: $d_{x^2-y^2}$, $d_{z^2}$). However, it is widely believed \cite{Daghofer2010,Luo2010} that the $t_{2g}$-orbitals are the most relevant orbitals close to Fermi surface. Furthermore, the $d_{yz}$ and $d_{xz}$ orbitals are often (or are close to be) degenerate and, as a consequence, one can design \cite{Herbrych2019} two-orbital models. Such choice represent a generic case of coexisting wide and narrow electronic bands, as often found in iron-based materials from the 123 family \cite{Medici2009,Luo2013,Rincon2014,Yi2017,Patel2019}. The particular choice of the hopping matrix elements $t_{\gamma\gamma^\prime}$ used here -- specifically $t_{00}=-0.5\,\mathrm{[eV]}$, $t_{11}=-0.15\,\mathrm{[eV]}$, and $t_{01}=t_{10}=0$ -- and crystal-field splitting $\Delta=0.8\,\mathrm{[eV]}$ is motivated by several previous studies \cite{Rincon2014,Herbrych2018,Herbrych2019} on the magnetic properties of the OSMP. The above values yield a kinetic energy bandwidth $W=2.1\,\mathrm{eV}$ which is used as an energy unit throughout the paper. Finally, in order to reduce the number of parameters of the Hamiltonian the value of Hund exchange will be fixed to $J_\mathrm{H}=U/4$ throughout our investigation. The rationale behind this value comes from dynamical mean field theory (using local-density approximation) calculations and is believed to be experimentally relevant \cite{Haule2009,Yin2011,Ferber2012}. Also, it was shown \cite{Zhang2012} that in a wide range of Hund couplings, the OSMP properties are the same \cite{Zhang2012,Rincon2014}.

Furthermore, it was previously shown \cite{Herbrych2019} that the magnetic properties of the OSMP can be qualitatively, and even quantitatively, described by the effective Hamiltonian obtained by Schrieffer-Wolff transforming the subspace with strictly one electron per site at the localized orbital $\gamma=1$, leading to the generalized Kondo-Heisenberg (gKH) model
\begin{eqnarray}
H_{\mathrm{K}}&=&-t_{\mathrm{00}}\sum_{\ell,\sigma}
\left(c^{\dagger}_{0,\ell,\sigma}c^{\phantom{\dagger}}_{0,\ell+1,\sigma}+\mathrm{H.c.}\right)
+U\sum_{\ell}n_{0,\ell,\uparrow}n_{0,\ell,\downarrow}\nonumber\\
&+&K\sum_{\ell}\mathbf{S}_{1,\ell} \cdot \mathbf{S}_{1,\ell+1}
-2J_{\mathrm{H}}\sum_{\ell}\mathbf{S}_{0,\ell} \cdot \mathbf{S}_{1,\ell}\,,
\label{hamkon}
\end{eqnarray}
where $K=4t_{11}^2/U$. It is worth noting that due to particle-hole symmetry, the electronic filling relation between the two--orbital Hubbard ($n_{\mathrm{H}}$) and gKH ($n_{\mathrm{K}}$) models is $n_{\mathrm{K}}=3-n_\mathrm{H}$.

In this work, we primarily reach our conclusions on the base of the gKH Hamiltonian, \eqref{hamkon}. However, in the {\it SI~Appendix} we reproduce the main findings with the full two--orbital Hubbard model \eqref{hamhub}. All Hamiltonians are diagonalized via the single-site density matrix renormalization group (DMRG) method~\cite{white1992,schollwock2005,white2005,Alvarez2009} (with up to $1200$ states kept), where the dynamical correlation functions are obtained using the dynamical-DMRG technique~\cite{jeckelmann2002,benthein2007,nocera2016}, i.e. calculating spectral functions directly in frequency space with the correction-vector method~\cite{kuhner1999} and Krylov decomposition~\cite{nocera2016}. Open boundary conditions are assumed.

%--------------------------------------------------------------------------------
\begin{figure}[!htb]
\includegraphics[width=1.0\columnwidth]{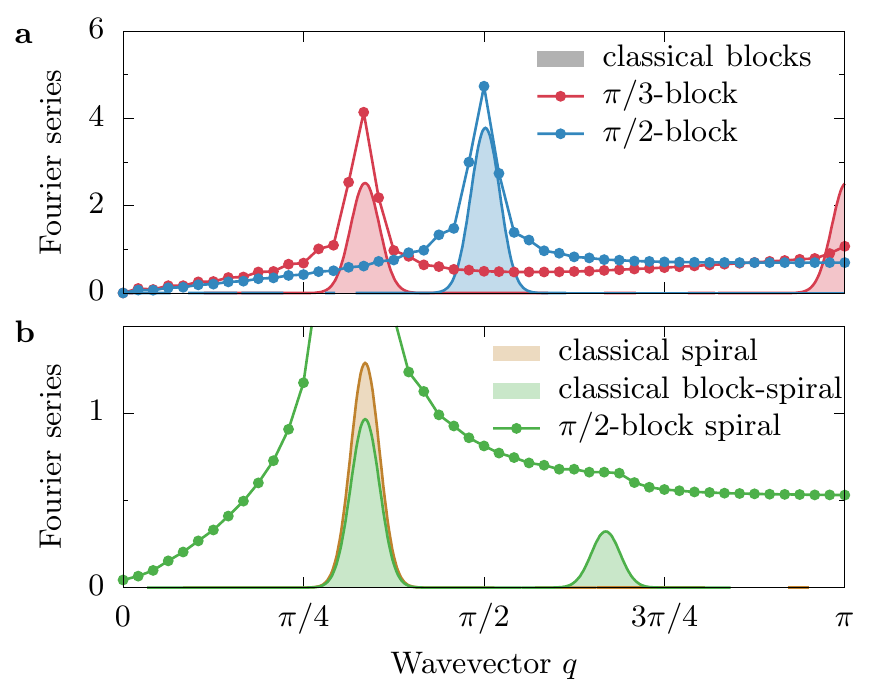}
\caption{Fourier decomposition of the spin order. {\bf (a)} Analysis of classical Heaviside-like block patterns (shaded area), i.e. $\uparrow\uparrow\downarrow\downarrow$ and $\uparrow\uparrow\uparrow\downarrow\downarrow\downarrow$. Line-dots represent the corresponding calculations using the generalized Kondo-Heisenberg model at $U=W$, $n_{\mathrm{H}}=1/2$ and $n_{\mathrm{H}}=1/3$ ($\pi/2$- and $\pi/3$-blocks, respectively). {\bf (b)} Shaded color areas represent (i) perfect standard spiral with one Fourier mode and (ii) our perfect block-spiral spin pattern displaying two modes. Line-dots are DMRG results for the block-spiral order at $U/W=1.95$ and $n_{\mathrm{H}}=0.5$. All results have pitch angle $\theta\simeq1/3$. The shoulder in the DMRG data is the {\it fingerprint} of the block-spiral. Note that in all panels we have broaden the the $\delta$-peaks of classical solutions for clarity.}
\label{classpin}
\end{figure}
%--------------------------------------------------------------------------------

Let us briefly describe the several magnetic phases of the OSMP, where the novel phase reported here is in yellow in Fig.~\ref{schematic}(b). For details we refer the interested reader to Ref.~\cite{Herbrych2019}. In Fig.~\ref{schematic}(b) we present a sketch of the interaction-filling ($U$--$n_{\mathrm{H/K}}$) phase diagram: (i) At $U<W$ the ground state is a paramagnetic metal. (ii) For $U\gtrsim W$ the system enters OSMP with coexisting metallic and Mott-insulating bands. (iii) For sufficiently large values of interaction $U\gg W$ the system is in a FM state for all fillings. (iv) When $U\sim {\cal O}(W)$, namely when all energy scales compete, the system is primarily in the so-called block-magnetic state. Depending on the filling of the itinerant band, the spins form various sizes of AFM-coupled FM spin islands. An important special case, found experimentally in BaFe$_2$Se$_3$, is the $n_{\mathrm{K}}=1/2$ ($n_{\mathrm{H}}=3/2+1$) filling where spins form a $\uparrow\uparrow\downarrow\downarrow\uparrow\uparrow\downarrow\downarrow$--pattern along the legs, the $\pi/2$-block magnetic state. 

The Fourier analysis of the {\it perfect} step-function pattern of the form $\uparrow\uparrow\downarrow\downarrow\uparrow\uparrow\downarrow\downarrow$ yields only one Fourier mode at $\pi/2$, as shown in Fig.~\ref{classpin}(a). Our explicit DMRG calculations of the gKH spin structure factor \mbox{$S(q)=\langle \mathbf{S}_q\cdot\mathbf{S}_{-q}\rangle$} [$\mathbf{S}_q=(1/\sqrt{L})\sum_\ell \mathrm{exp}(-iq\ell) \mathbf{S}_{\ell}$ with $\mathbf{S}_{\ell}=\sum_{\gamma}\mathbf{S}_{\gamma,\ell}$] at $U\sim W$ confirms that the dominant contribution to the magnetic ordering indeed originates in a $\pi/2$-block pattern. Equivalently, the analysis of a perfect $\uparrow\uparrow\uparrow\downarrow\downarrow\downarrow$ pattern yields now {\it two} equal-height Fourier components at $\pi/3$ and $\pi$. Calculations within gKH [see Fig.~\ref{classpin}(a)] display a large, dominant peak at $\pi/3$ but also a smaller one at $\pi$. The small weight of the latter can be explained by the emergence of optical modes of localized spin excitations present within multi-orbital systems \cite{Herbrych2018}. These modes manifest in $S(q)$ as a finite offset at large values of the wavevector [see Fig.~\ref{schematic}(c)]. Nevertheless, the dominant shape of the block magnetism of OSMP can be qualitatively described by idealized Heaviside-like patterns. The small size of our blocks show that they cannot be charactized as domain walls, that are usually separated by much larger distances, but as a new magnetic order.

%--------------------------------------------------------------------------------
\begin{figure*}[!htb]
\includegraphics[width=1.0\textwidth]{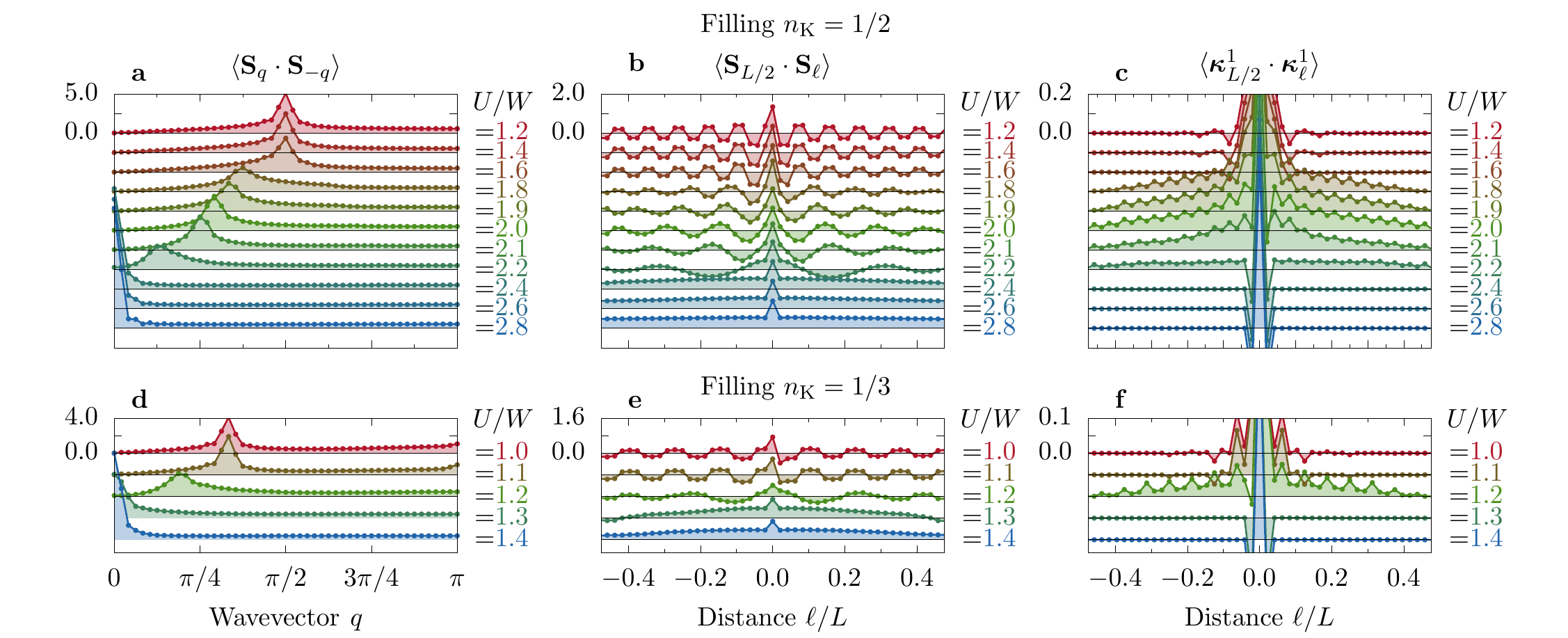}
\caption{Interaction $U/W$ dependence of correlation functions. ({\bf a,d}) Static spin structure factor $\langle \mathbf{S}_q\cdot\mathbf{S}_{-q}\rangle $. ({\bf b,e}) Real-space spin-spin correlation function $\langle \mathbf{S}_{L/2}\cdot \mathbf{S}_\ell\rangle$, ({\bf c,f}) nearest--neighbour chirality correlation function $\langle \boldsymbol{\kappa}^1_{L/2}\cdot\boldsymbol{\kappa}^1_\ell\rangle$. Top row [panels ({\bf a-c})] represents results for filling $n_{\mathrm{K}}=1/2$, while bottom row [panels ({\bf d-f})] depicts results for $n_{\mathrm{K}}=1/3$. All results calculated using DMRG and the generalized Kondo-Heisenberg model on $L=48$ sites.}
\label{realspace}
\end{figure*}
%--------------------------------------------------------------------------------

%--------------------------------------------------------------------------------
\vspace{1em}
{\bf B. Chirality correlations.} In between the block and FM phases, we discovered a novel region where the spin static structure factor $S(q)$ have its maximum $q_{\mathrm{max}}$ at incommensurate values of the wavevector $q$. In Figs.~\ref{realspace}(a) and (d) we explicitly show the interaction $U$ dependence of $S(q)$ in this region (see also Ref.~\cite{Herbrych2019}) for $n_{\mathrm{K}}=1/2$ and $n_{\mathrm{K}}=1/3$, respectively. It is evident from the presented results that the maximum of the spin structure factor continuously changes with interaction $U$ interpolating between block-magnetism at $U\sim W$ and ferromagnetic (FM) state at $U\gg W$ (e.g., at fixed $n_{\mathrm{K}}=1/2$, from $q_{\mathrm{max}}=\pi/2$ to $q_{\mathrm{max}}=0$). The real-space spin-spin correlation functions $\langle \mathbf{S}_{L/2}\cdot \mathbf{S}_{\ell}\rangle$ [see Figs.~\ref{realspace}(b,e)] reveal an oscillatory structure throughout the chain, with period $\theta=q_{\mathrm{max}}$, decaying in amplitude at large spatial separations, within the incommensurate region in the phase diagram. Such behaviour may naively suggest a spin-density wave. For the latter, the Fourier transform of the spin correlations should yield only one Fourier mode, due to $\langle \mathbf{S}_{L/2}\cdot \mathbf{S}_{\ell}\rangle\propto\cos(q_{\mathrm{max}}\ell)$. As we will show, this interpretation is incorrect and the competing interactions present in OSMP systems, as well as the location of this phase sandwiched between block and FM states, lead to a novel type of spiral state. 

To better investigate the magnetic structure of the novel spiral within the OSMP we will focus on the {\it chirality} correlation function $\langle \boldsymbol{\kappa}_\ell^d\cdot\boldsymbol{\kappa}_m^d\rangle$ \cite{Hikihara2000,Hikihara2001,McCulloch2008,Sato2011}, where
\begin{equation}
\boldsymbol{\kappa}_\ell^d=\mathbf{S}_{\ell}\times \mathbf{S}_{\ell+d}\,,
\end{equation}
represents the vector product of two spin operators separated by a distance $d$, and consequently the angle between them. We stress that in the following we consider the total spin at each site $\mathbf{S}_{\ell}=\sum_\gamma \mathbf{S}_{\gamma,\ell}$. In the generic case of AFM-- or FM--order, and also in the OSMP block--phase, the chirality correlation function vanishes, $\langle \boldsymbol{\kappa}_\ell^d\cdot\boldsymbol{\kappa}_m^d\rangle=0$, since the spins are collinear. On the other hand, consecutive $\langle \boldsymbol{\kappa}_\ell^d\cdot\boldsymbol{\kappa}_m^d\rangle\ne0$ indicate a nontrivial spiral order. 

\begin{figure}[!htb]
\includegraphics[width=1.0\columnwidth]{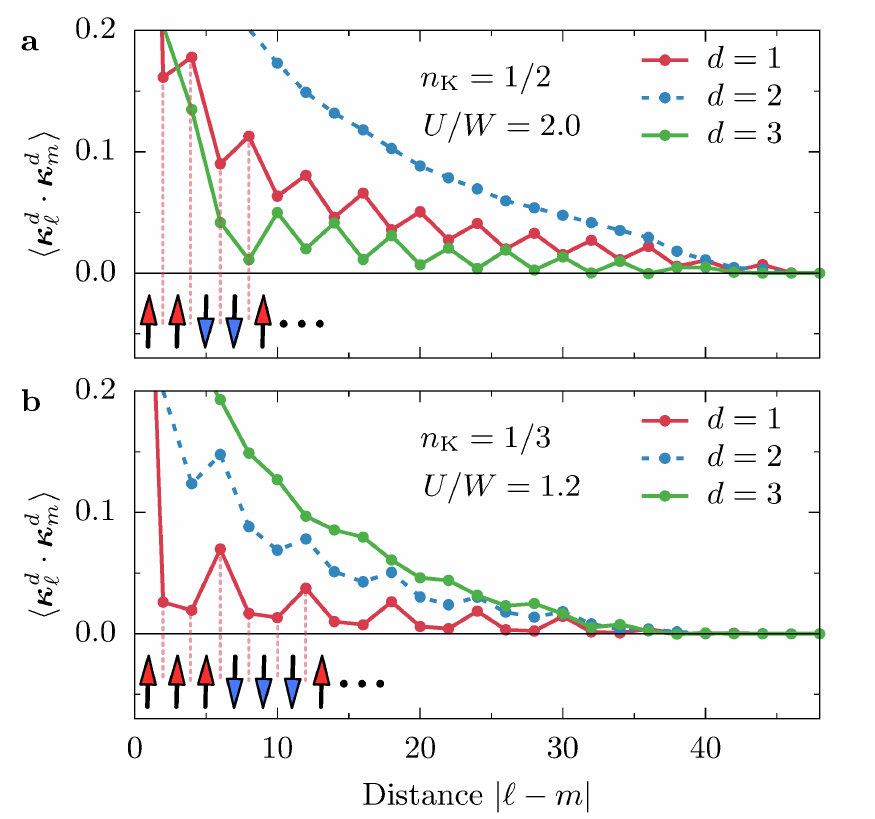}
\caption{Chiral correlation function $\langle \boldsymbol{\kappa}^d_\ell \cdot \boldsymbol{\kappa}^d_m\rangle$. ({\bf a}) Dependence of chirality $\boldsymbol{\kappa}^d_\ell=\mathbf{S}_\ell\times\mathbf{S}_{\ell+d}$ on distance $d$ between spins. Results calculated using the generalized Kondo-Heisenberg, $L=48$, $n_{\mathrm{K}}=1/2$, and $U/W=2.0$. Panel ({\bf b}): same as ({\bf a}) but for $n_{\mathrm{K}}=1/3$ and $U/W=1.2$. In both panels: arrows represent schematics of the block order for given filling.}
\label{kappa}
\end{figure}

In Fig.~\ref{realspace}(c,f) we present the interaction $U$ dependence of the nearest--neighbour, $d=1$, chirality correlation function $\langle \boldsymbol{\kappa}_\ell^1\cdot\boldsymbol{\kappa}_m^1\rangle$ of the gKH model. It is evident from the presented results that between the block and FM phases the chirality acquires finite values. In addition, the spatial structure of $\langle \boldsymbol{\kappa}_\ell^1\cdot\boldsymbol{\kappa}_m^1\rangle$ displays a clear {\it zig-zag}--like pattern. To better investigate the spiral internal structure, in Fig.~\ref{kappa} we present the dependence of the chirality correlation with the distance $d$ between spins. In {\it SI~Appendix} we provide the full interaction $U$ dependence of the next--nearest--neighbour, $d=2$, chirality correlation function. Here, as illustration we will focus on the representative cases of $U/W=2.0$ for $n_{\mathrm{K}}=1/2$ and $U/W=1.2$ for $n_{\mathrm{K}}=1/3$. 

%--------------------------------------------------------------------------------
%--------------------------------------------------------------------------------

As already mentioned, the nearest--neighbour ($d=1$) chiral correlation for both considered fillings contains additional patterns modulating the usual decay. Specifically, for $n_{\mathrm{K}}=1/2$ ($n_{\mathrm{K}}=1/3$) the correlation function {\it oscillates} every two (three) sites. Interestingly, these patterns change their nature when the next--nearest neighbour ($d=2$) chirality is considered: (i) the values of these chiralities increase $\langle \boldsymbol{\kappa}_\ell^2\cdot\boldsymbol{\kappa}_m^2\rangle>\langle \boldsymbol{\kappa}_\ell^1\cdot\boldsymbol{\kappa}_m^1\rangle$, and (ii) for the case of $n_{\mathrm{K}}=1/2$ the $\boldsymbol{\kappa}$-correlation is now a smooth function of distance, while $n_{\mathrm{K}}=1/3$ still exhibits some three-site oscillations. Investigating the next-next--nearest neighbour case, $d=3$, gives additional information. While for the $n_{\mathrm{K}}=1/2$ filling the $d=3$ correlations are smaller than $d=1$ and $d=2$, for $n_{\mathrm{K}}=1/3$ they are larger and (as for $d=2$ at $n_{\mathrm{K}}=1/2$) they are now a smooth function of distance. This seemingly erratic behaviour of $\langle\boldsymbol{\kappa}^d_\ell\cdot\boldsymbol{\kappa}^d_m\rangle$ correlations varying $d$ cannot be simply explained by a mere uniform change of the pitch angle $\theta$. The latter changes only with the interaction $\theta=\theta(U)$. On the other hand, as evident from Figs.~\ref{realspace}(c,f), the internal structure of $\langle \boldsymbol{\kappa}_\ell^1\cdot\boldsymbol{\kappa}_m^1\rangle$ depends only on the electronic filling $n_{\mathrm{K}}$ (see also {\it SI~Appendix} for $d=2$ results). 

%--------------------------------------------------------------------------------
\begin{figure}[!htb]
\includegraphics[width=1.0\columnwidth]{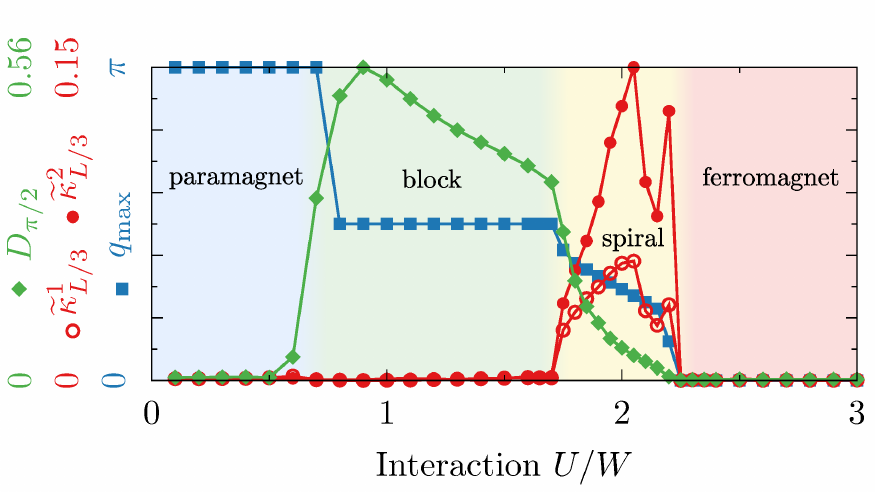}
\caption{Phase diagram varying the interaction $U/W$. Presented are: (i) maximum of static spin structure factor $q_{\mathrm{max}}$ (squares), (ii) nearest-- ($d=1$) and next--nearest ($d=2$) neighbour chirality $\widetilde{\kappa}^{d}_{L/3}$ (open and filled circles, respectively), and (iii) dimer correlation $D_{\pi/2}$ (diamonds). Results calculated using the generalized Kondo-Heisenberg model, $L=48$ sites, and $n_\mathrm{K}=1/2$.}
\label{phase}
\end{figure}
%--------------------------------------------------------------------------------

To better explain this behaviour let us focus on the dimer correlation defined \cite{Xavier2003} as
\begin{equation}
D_{\pi/2}=\frac{2}{L}\sum_{\ell=L/4}^{3L/4} (-1)^{\ell-1} \langle \mathbf{S}_{\ell}\cdot \mathbf{S}_{\ell+1}\rangle\,.
\end{equation}
The above operator {\it compares} the number of FM and AFM bonds in the bulk system ({\it SI~Appendix} contains the full real-space dependence of $\langle \mathbf{S}_{\ell}\cdot \mathbf{S}_{\ell+1}\rangle$). For true FM or AFM ordered states, $|\uparrow\uparrow\uparrow\uparrow\dots\rangle$ or $|\uparrow\downarrow\uparrow\downarrow\dots\rangle$ respectively, each nearest--neighbour bond has the same sign: positive for FM $\uparrow\uparrow$ and negative for AFM $\uparrow\downarrow$. Consequently, $D_{\pi/2}=0$. On the other hand, in the $\pi/2$-block state, $|\uparrow\uparrow\downarrow\downarrow\dots\rangle$, the FM and AFM bonds alter in staggered fashion rendering $D_{\pi/2}\ne0$. In Fig.~\ref{phase} we present the interaction $U$ dependence of $D_{\pi/2}$ for $n_{\mathrm{K}}=1/2$. Furthermore, in the same figure we present also the wavevector where the static structure factor is maximized $q_{\mathrm{max}}$, and the value of the $\widetilde{\kappa}_{L/3}^d=\langle \boldsymbol{\kappa}^d_{\ell}\cdot\boldsymbol{\kappa}^d_{m}\rangle_{|\ell-m|=L/3}$ correlator for $d=1$ and $d=2$. Starting in the paramagnet at small $U$ both $D_{\pi/2}$ and $\widetilde{\kappa}_{L/3}^d$ vanish, with $q_{\mathrm{max}}=\pi$ just depicting the usual short-range staggered correlations of weak-$U$ physics. In the opposite limit of strong interaction, $U\gg W$, $D_{\pi/2}=\widetilde{\kappa}_{L/3}^d=0$ as well, consequence of a {\it simple} FM-state with $q_{\mathrm{max}}=0$. In the most interesting case of competing interaction $U\sim W$, the dimer correlation $D_{\pi/2}$ acquires a finite value maximized at $U\simeq W$. The latter reflects a {\it perfect} $\pi/2$-block magnetic state. Interestingly, one can observe a continuous transition of $D_{\pi/2}$ between the block-- and FM--phases in the region where a finite chirality $\widetilde{\kappa}_{L/3}^d\ne0$ was found and where $q_{\mathrm{max}}$ takes incommensurate values.

On the basis of the above results, a coherent picture emerges explaining the nature of the magnetic state between the block and FM limits. Consider first filling $n_{\mathrm{K}}=1/2$. At $U\simeq W$ the ground state is a block--magnetic phase, where two-site FM islands (blocks) are AFM coupled. Increasing the interaction $U$, the spins start to rotate w.r.t. each other, inducing finite $\langle \boldsymbol{\kappa}^d_{\ell}\cdot\boldsymbol{\kappa}^d_{m}\rangle$ correlations. Remarkably, during the rotation the overall {\it FM islands}--nature of the state is preserved, yielding a finite $D_{\pi/2}\ne0$ all the way to the FM--state at $U\gg W$. Such an unexpected scenario is also encoded in the inequalities $\widetilde{\kappa}_{L/3}^2>\widetilde{\kappa}_{L/3}^1$ and $\widetilde{\kappa}_{L/3}^2>\widetilde{\kappa}_{L/3}^3$ observed in Fig.~\ref{kappa}(a). This is qualitatively different from a standard spiral state where the spin rotates from site to site [top sketch in Fig.~\ref{schematic}(a)]. In our case, instead, the spiral is made of individual blocks, and it is the entire block that rotates from block-to-block [middle sketch in Fig.~\ref{schematic}(a)]. Furthermore, the detail analysis of $S(q)$ reveals a small secondary Fourier mode at $\pi-q_{\mathrm{max}}$ [see Fig.~\ref{classpin}(b)]. As already mentioned, the long-wavelength components of $S(q)$ are hidden behind the OSMP optical mode contribution \cite{Herbrych2018} and a detailed analysis of experimental dynamical spin spectra will be needed to fully reveal the presence of our predicted block-spiral states. Because this additional modes are not consistent with a mere standard spiral but instead appears in the Fourier analysis of the perfectly sharp block $\uparrow\uparrow\downarrow\downarrow$ state modulated by the spiral $\cos$-like component, and are also consistent with our analysis of $\widetilde{\kappa}$, they represent the {\it fingerprints} of our novel block-spiral states.

%--------------------------------------------------------------------------------
\begin{figure*}[!htb]
\includegraphics[width=0.9\textwidth]{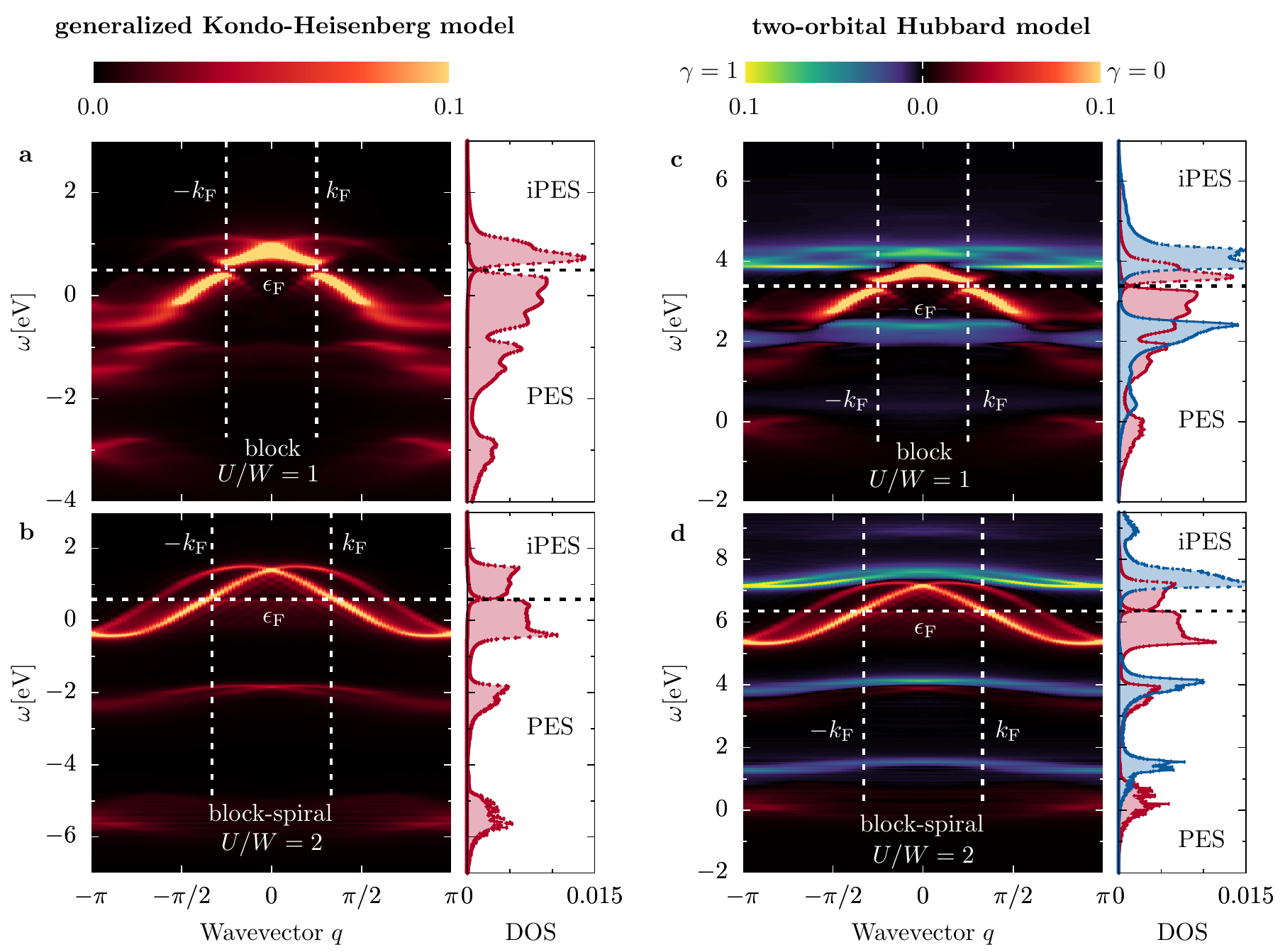}
\caption{Spectral function $A(q,\omega)$. ({\bf a}) Block-phase ($n_{\mathrm{K}}=3/2\,,U/W=1.0$) and ({\bf b}) block-spiral-phase ($n_{\mathrm{K}}=3/2\,,U/W=2.0$) as calculated for the generalized Kondo-Heisenberg model using $L=48$ sites. ({\bf c},{\bf d}) The same results obtained with the two-orbital Hubbard model ($L=48$). ({\bf c}) Block-phase ($n_{\mathrm{H}}=2.50\,,U/W=1.0$) and ({\bf d}) block-spiral-phase ($n_{\mathrm{H}}=2.50\,,U/W=2.0$). Horizontal (vertical) lines depict the Fermi level $\epsilon_{\mathrm{F}}$ (Fermi wavevector $k_{\mathrm{F}}$), while (i)PES stands for (inverse--) photoemission spectroscopy. Right panels: density of states $\mathrm{DOS}(\omega)=\sum_q A(q,\omega)$.}
\label{akw}
\end{figure*}
%--------------------------------------------------------------------------------

Such a novel block-spiral state can also be observed at filling $n_{\mathrm{K}}=1/3$. Here, $\pi/3$--blocks of three sites $\uparrow\uparrow\uparrow\downarrow\downarrow\downarrow$ develop a finite dimer correlation of the form $D_{\pi/3}\propto \sum_\ell f(\ell)\langle \mathbf{S}_{\ell}\cdot \mathbf{S}_{\ell+1}\rangle$, where $f(\ell)$ accounts for the specific form of the bond sign pattern, i.e. $\{1,1,-1,1,1,-1,\dots\}$ (see {\it SI~Appendix}). A finite $D_{\pi/3}$ together with $\widetilde{\kappa}_{L/3}^3>\widetilde{\kappa}_{L/3}^2>\widetilde{\kappa}_{L/3}^1$ is compatible with a spiral state of rotating three-site blocks [see bottom sketch in Fig.~\ref{schematic}(a)].

Finally, let us comment on the finite-size dependence of our findings. Analysis of system sizes up to $L=96$ sites (see {\it SI~Appendix}) indicates that the discussed block-spiral states display short-ranged order but with a robust correlation length of $\xi\sim15$ sites, where $\langle \boldsymbol{\kappa}^d_{\ell}\cdot\boldsymbol{\kappa}^d_{\ell+x}\rangle\propto \exp(-x/\xi)$. However, it was argued \cite{Sato2011,Maghrebi2017} for the case of the FM long-range Heisenberg model that realistic small SU(2)-breaking anisotropies, often present in real materials, can induce true (quasi-)long-range order in a spiral state. Also, such an anisotropy will {\it choose} the plane of rotation of the spiral, i.e. in-plane or out-of-plane with regards to the chain direction. As we will argue in the next section, (frustrated) long-range Heisenberg Hamiltonians can (at least qualitatively) capture the main physics of the block-spiral unveiled here.

We emphasize that the same conclusions are reached in the multi-orbital Hubbard model, although because the effort is much more computationally demanding it was limited to special cases. In {\it SI~Appendix} we present results for $\langle \boldsymbol{\kappa}^d_\ell\cdot\boldsymbol{\kappa}^d_m\rangle$ obtained with the full two--orbital Hubbard model \eqref{hamhub} and also the incommensurability of $S(q)$ for a three-orbital Hubbard model.

\vspace{1em}
{\bf C. Quasi-particle excitations.} A distinctive feature of the OSMP is the coexistence of localized electrons (spins in an insulating band) and itinerant electrons (a metallic band). In the block-magnetic phase at $U\simeq W$ it was previously shown \cite{Patel2019} -- for the three-orbital Hubbard model -- that the density of states (DOS) at the Fermi level $\epsilon_{\mathrm{F}}$ is reduced, indicating a pseudogap-like behaviour. Our calculations of the single-particle spectral function $A(q,\omega)$ and $\mathrm{DOS}(\omega)$ (see {\it SI~Appendix}) using the gKH model are presented in Fig.~\ref{akw}(a) and confirm this picture. They also show the strength of our effective Hamiltonian: the behaviour of the gKH model perfectly matches the $\gamma=0$ itinerant orbital of the full two-orbital Hubbard result presented in Fig.~\ref{akw}(c). It is worth noticing that to properly match the electron and hole parts of $A(k,\omega)$ between the models we exploit the particle-hole symmetry of gKH and present results for $n_{\mathrm{K}}=3/2$ (instead of $n_{\mathrm{K}}=1/2$ for which the spectrum would be simply {\it mirrored}, i.e., $\omega\to -\omega$ and $k\to \pi-k$). Also, we want to reiterate here that although the system is overall metallic in nature, the band structure is vastly different from the simple cosine-like result of $U\to 0$. Distinctive features in $A(q,\omega)$ at the Fermi vector $k_{\mathrm{F}}$, and a large renormalization of the overall band structure at higher energies, indicate a complex interplay between various degrees of freedom and energy scales. 

Upon increasing the interaction $U$ and entering the block-spiral region, $A(q,\omega)$ changes drastically. In Fig.~\ref{akw}(b) we show representative results for a $\theta/\pi\simeq 0.3$ block-spiral state at $U/W=2$ and $n_{\mathrm{K}}=1/2$. Two conclusions are directly evident from the presented results: (1) the pseudogap at $\epsilon_{\mathrm{F}}$ is closed, but some additional gaps at higher energies opened. (2) $A(q,\omega)$ in the vicinity of the Fermi level, $\omega\sim\epsilon_{\mathrm{F}}$, develops two bands, intersecting at the $q=0$ and $q=\pi$ points, with maximum at $q\simeq\theta/2$. The bands represents two {\it quasiparticles}: {\it left}- and {\it right}-movers reflecting the two possible rotations of the spirals. It is obvious from the above results that the quasiparticles break the parity symmetry, i.e., going from $q\to -q$ momentum changes the quasiparticle character, as expected for a spiral state. Somewhat surprisingly, $A(q,\omega)$ does not show any gap as would typically be associated with the finite dimerization $D_{\alpha}$ that we observe. However, it should be noted that for quantum localized $S=1/2$ spins the quarter-filling ($n_{\mathrm{K}}=1/2$) implies a filling of 2/5 of the lower Kondo band (due to energy difference between local Kondo singlets and triplets). The dimerization gap expected at $\pi/2$ would thus open away from the Fermi level and would consequently not confer substantial energy gain to the electrons. Similarly, no dimerization gap is found in the $n_{\mathrm{K}}\simeq 0.33$ case, i.e., the $\pi/3$-block spiral (not presented). We thus conclude that quantum fluctuations of the localized and itinerant spins are here strong enough to suppress the dimerization gap. The above conclusions can be reached from results obtained with the full two-orbital Hubbard model [see Figs.~\ref{akw}(d)].

%--------------------------------------------------------------------------------
\vspace{1em}
\noindent {\bf \large Discussion and effective model}
\vspace{1em}
%--------------------------------------------------------------------------------

%--------------------------------------------------------------------------------
\begin{figure}[!htb]
\includegraphics[width=1.0\columnwidth]{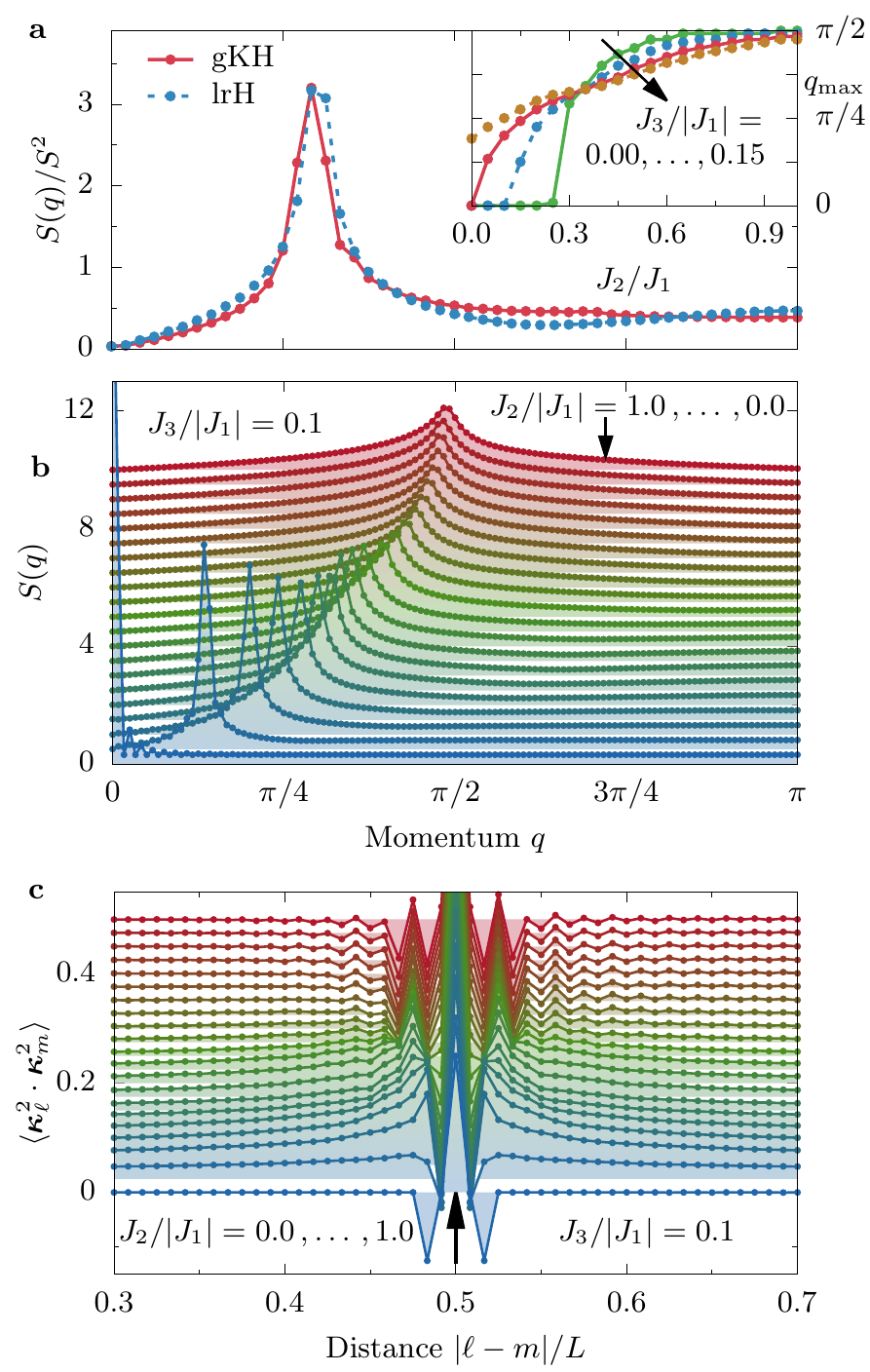}
\caption{Effective Heisenberg model. ({\bf a}) Comparison of $S(q)$ between the generalized Kondo-Heisenberg (gKH) model in the spiral state ($L=48\,,n_{\mathrm{K}}=1/2\,,U/W=2.0$) and the long-range Heisenberg (lrH) model ($L=48\,,J_2/|J_1|=0.25\,,J_3/|J_1|=0.1$). Note results are normalized by the magnetic moment $S^2=\mathbf{S}(\mathbf{S}+1)$, i.e., $S^2=1.375$ and $S^2=3/4$ for gKH and lrH, respectively. ({\bf b}) Static spin structure factor $S(q)$ and ({\bf c}) chirality correlation function $\langle \boldsymbol{\kappa}^d_\ell\cdot\boldsymbol{\kappa}^d_m\rangle$ of the $J_1$-$J_2$-$J_3$ Heisenberg model (with ferromagnetic $J_1=-1$), calculated for $L=120$ sites, $J_3/|J_1|=0.1$, and various $J_2/|J_1|=0.00\,,0.05\,,\dots,1.00$ (top to bottom). Inset of ({\bf a}): position of the maximum of the static structure factor $q_{\mathrm{max}}$ of the lrH model as function of $J_2/|J_1|$, for $J_3/|J_1|=0.00\,,0.05,0.10$, and $0.15$.}
\label{spin}
\end{figure}
%--------------------------------------------------------------------------------

It was previously shown \cite{Herbrych2018} that the frustrated FM-AFM $J_1$-$J_2$ Heisenberg model (with $|J_1|\sim J_2$) qualitatively captures the physics of the non-spiral block-magnetic state. Here, in Fig.~\ref{spin}(a) we show that the spin structure factor $S(q)$ of the block-spiral state can be accurately described by an extension of that model: the frustrated {\it long-range} Heisenberg Hamiltonian. Although this {\it phenomenological} model is not derived here from the basic Hamiltonians, it accurately reproduces the interaction $U$ dependence throughout the incommensurate region [e.g. compare Fig.~\ref{realspace}(a) and Fig.~\ref{spin}(b)]. The intuitive understanding of the origin of the effective spin model is as follows: at $U\sim W$ the system is in a block-magnetic state with quasi-long-range (QLR) spin correlation of $\pi/N_{\mathrm{b}}$-nature, where $N_\mathrm{b}$ is the size of the block. We found excellent agreement between the gKH and Heisenberg models with $J_1=-1$, $J_2=1/4$, $J_3=1/10$ (see also inset of Fig.~\ref{spin} for the $J_2$-$J_3$ dependence of $q_{\mathrm{max}}$). The later yields $\alpha\simeq 2$ in $|J_r|\propto1/r^\alpha$. As a consequence, the block-spiral magnetism is described by the class of Haldane-Shastry models \cite{Haldane1988,Shastry1988}. In fact, we speculate that due to the oscillating character of the electron mediated exchange couplings, the spin system which encompasses all phenomena is given by
\begin{eqnarray}
H&=&\sum_{\ell,r}\,J_r\,\mathbf{S}_\ell\cdot \mathbf{S}_{\ell+r}\,,\nonumber\\
\text{with}\quad J_1<0\,,&\quad& J_2>0\,,\quad J_{r>2}=|J_1|\frac{(-1)^{r-1}}{r^\alpha}\,.
\label{hamheis}
\end{eqnarray}
It was shown \cite{Sandvik2004,Laflorencie2005,Li2015} that the above model have (for the zero magnetization sector, $S^z_{\mathrm{tot}}=0$) a QLR ground state with $\pi/2$-correlations and also can support spiral states. Such frustrated long-range Hamiltonians are not suitable for direct DMRG calculations due to the area law of entanglement. Thus, alternatively in {\it SI~Appendix} we show small system size Lanczos diagonalization results for \eqref{hamheis}. However, as presented in Fig.~\ref{spin}, the first three terms $J_1$-$J_2$-$J_3$ (which can still be computed accurately with DMRG) already give satisfactory results even for the chirality correlator [compare Fig.~\ref{realspace}(c) and Fig.~\ref{spin}(c)].

Finally, two additional comments: (1) the scenario (long-range effective spin model) described above goes beyond the discussed non-spiral $\pi/2$-block case. Changes in the magnetization sector of the long-range FM Hamiltonian lead to modifications in the periodicity of the QLR order \cite{Ueda2014,Onishi2015}. As a consequence, there are only a few parameters in the effective spin model: the magnetization $S^z_{\mathrm{tot}}$ related to the filling in the original full multi-orbital Hamiltonian and $\{J_2,\alpha\}=f(U)$ which controls the long-range nature of the system. (2) Since our system is overall metallic (albeit likely a bad metal because of the localized component), one could naively believe that the Ruderman-Kittel-Kasuya-Yosida (RKKY) spin exchange $J_r$ carried by mobile electrons could explain some of our results. However, this is not the case because the nontrivial effect of the interaction $U$ -- creating a metallic state coexisting with a (quasi)-ordered magnetic state -- qualitatively modifies the nature of the RKKY interaction. 

%--------------------------------------------------------------------------------
\vspace{1em}
\noindent {\bf \large Conclusion}
\vspace{1em}
%--------------------------------------------------------------------------------

In conclusion, we have identified a novel type of spiral spin order: block-spiral magnetism in which FM islands rigidly {\it rotate} with respect to each other. We wish to emphasize the crucial role of correlation in this phenomenon. Spiral states are usually a consequence of frustration (geometrical or induced by competing interactions) or by explicit symmetry breaking terms like DM couplings. For example, long period helical spin density waves where reported \cite{Bak1982,Kataoka1987,Jensen1991,Mochizuki2012} in many transition-metal compounds and rare-earth magnets. This is also the case of the hexagonal perovskite CsCuCl$_3$, where the external magnetic field can induce block-like structure on top of the spiral-like order \cite{Stuber2004}. However, in all of these cases the magnetic structure resembles a domain wall and originates in strong frustration of the (often classical) model itself. On the contrary, in our system we have only nearest--neighbour interactions on a chain geometry, and the SU(2) symmetry is preserved. Instead, the novel block-spiral state reported here appears as an effect of {\it hidden frustration}, i.e., competition between the double-exchange like mechanism present in multi-orbital systems dominated by a robust Hund exchange and the interaction $U$ that governs superexchange tendencies. These nontrivial effects become apparent in the exotic effective (phenomenological) spin model we unveiled: a long-range frustrated Heisenberg model. Within the latter the spin exchange decays slowly with distance, $\sim1/r^2$, in contrast to the usual RKKY interaction which decays faster. Also, to our knowledge the multi-orbital system discussed in this work - as realized in iron-based compounds from the 123 family - could be one of the few, if not the only, known realization of a Haldane-Shastry--like model with $\alpha=2$.

Furthermore, due to properties unique to the OSMP, primarily the coexistence of metallic and insulating bands, the block-spiral state displays exotic behavior in the electronic degrees of freedom. For example, new quasiparticles appear due to the parity breaking of the spiral state (left- and right-movers). Similar physics can be found in systems with spin-orbit coupling \cite{Li2012,Winkler2017}. Here, again, this is an effect of competing energy scales. Another interesting possibility is the existence of multiferroic behaviour in our system. It is known \cite{Seki2008,Furukawa2010,Tokura2010,Povarov2016,Scaramucci2018} that in materials such as quasi-1D compounds LiCu$_2$O$_2$, LiCuVO$_2$ or PbCuSO$_4$(OH)$_2$, the spin spirals drive the system to ferroelectricity. Moreover, the phenomena described here, robust spiral magnetism without a charge gap, is at the heart of one of the proposed systems where topological Majorana phases can be induced \cite{Oreg2010,Braunecker2013,Klinovaja2013,Perge2014,Schecter2015,Steinbrecher2018}.

Finally, let us comment on our results from the perspective of the real iron-based materials. As already mentioned, a nontrivial magnetic order, such as spirals, in the vicinity of high critical temperature superconductivity can lead to topological effects. This is the case of the 2D material FeTe$_{1-x}$Se$_x$ where zero-energy vortex bound states (Majorana fermions) have been reported \cite{Zhang2018-1,Wang2018,Zhang2018-2,Machida2019,Zhang2019}. Furthermore, similar to our findings, it was argued \cite{Ruiz2019} that the frustrated magnetism of FeTe$_{1-x}$Se$_x$ can be captured by a long-range $J_1$-$J_2$-$J_3$ spin Hamiltonian. From this perspective, it seems appropriate to assume that the phenomena described in our work extends beyond 1D systems. Unfortunately, the lack of sufficiently reliable computational methods to treat 2D quantum models limits our understanding of multi-orbital effects in 2D. An intermediate promising route are the low-dimensional ladders from the family of 123 compounds where accurate DMRG calculations are possible. Early density functional theory and Hartree--Fock results suggest that the effects of correlations are important \cite{Luo2013,YZhang2019} and that noncollinear magnetic order can develop in the ground-state \cite{Luo2013}. The recent proposal of exploring ladder tellurides with a predicted higher value of $U/W$ provides another avenue to consider \cite{YZhang2019}. As a consequence, we encourage crystal growers with expertise in iron-based materials to explore in detail the low-dimensional family of 123 compounds, including doped samples, because our results suggest that new and exotic physics may come to light.

%================================================================================

%================================================================================

\vspace{2em}
\noindent {\bf \large Acknowledgments}\\
We thank M.~L.~Baez, C.~Batista, and M.~Mierzejewski for fruitful discussions. J.~Herbrych, A.~Moreo, and E.~Dagotto were supported by the US Department of Energy (DOE), Office of Science, Basic Energy Sciences (BES), Materials Sciences and Engineering Division. In addition, J.~Herbrych acknowledges grant support by the Polish National Agency of Academic Exchange (NAWA) under contract PPN/PPO/2018/1/00035. The development of the DMRG++ code by G. Alvarez was supported by the Scientific Discovery through Advanced Computing (SciDAC) program funded by the U.S. DOE, Office of Science, Advanced Scientific Computer Research and Basic Energy Sciences, Division of Materials Science and Engineering, which was conducted at the Center for Nanophase Materials Science, sponsored by the Scientific User Facilities Division, BES, DOE, under contract with University of Tennessee-Battelle. J.~Heverhagen and M.~Daghofer were supported by the Deutsche Forschungsgemeinschaft, via the Emmy-Noether program (DA 1235/1-1) and FOR1807 (DA 1235/5-1) and by the state of Baden-W\"{u}rttemberg through bwHPC. Calculations have been partly carried out using resources provided by Wroclaw Centre for Networking and Supercomputing.

%================================================================================
\clearpage
\appendix
\setcounter{figure}{0}
\setcounter{equation}{0}
\setcounter{page}{1}
\renewcommand{\thefigure}{S\arabic{figure}}
\renewcommand{\theequation}{S\arabic{equation}}
%================================================================================
\onecolumngrid
\begin{center}
{\bf \uppercase{Supplementary Information}} for:\\
\vspace{5pt}
{\bf \Large Block-spiral magnetism: An Exotic Type of Frustrated Order}\\
\vspace{5pt}
by J. Herbrych, J. Heverhagen, G. Alvarez, M. Daghofer, A. Moreo, E. Dagotto
\end{center}
\vspace{20pt}
\twocolumngrid

%================================================================================
\begin{center}
{\bf \uppercase{Supplementary Note 1}\\Chirality correlations}
\end{center}

\begin{figure}[!htb]
\includegraphics[width=0.5\textwidth]{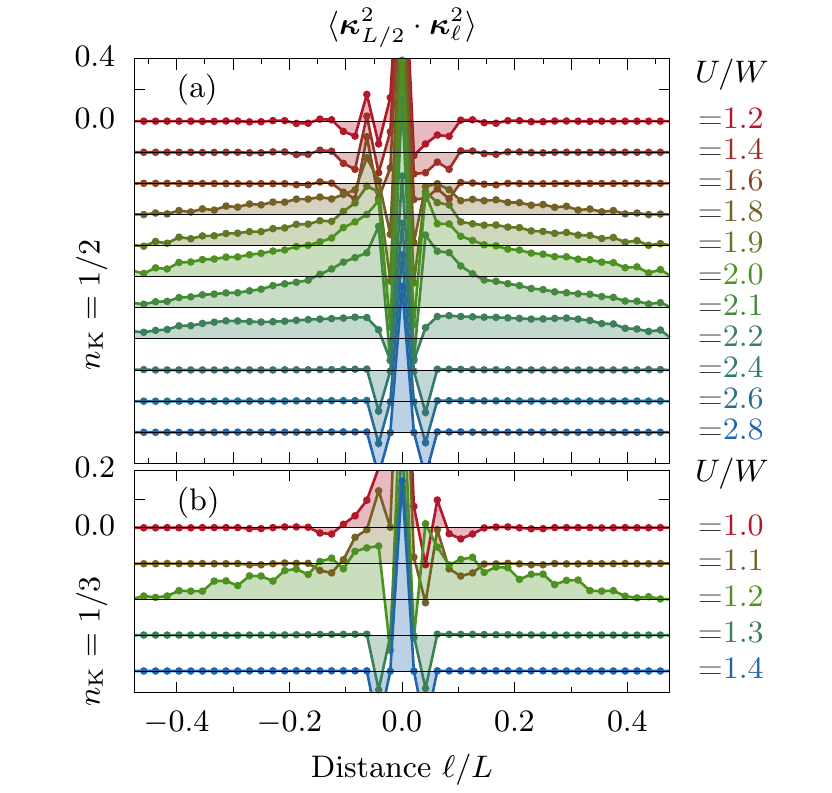}
\caption{Interaction $U$ dependence of the next--nearest neighbour, $d=2$, chirality correlation function $\langle \boldsymbol{\kappa}^2_{L/2}\cdot\boldsymbol{\kappa}^2_\ell\rangle$, using the generalized Kondo-Heisenberg model on $L=48$ sites. Panel (a) are results for filling $n_{\mathrm{K}}=1/2$, while panel (b) depicts results for $n_{\mathrm{K}}=1/3$.}
\label{realspace_s}
\end{figure}

In Fig.~\ref{realspace_s} we present data that is complementary to that shown in Fig.~3 of the main text for the next--nearest neighbour, $d=2$, chirality correlation function. These results are for the generalized Kondo-Heisenberg model calculated for $n_{\mathrm{K}}=1/2$ [panel (a)] and $n_{\mathrm{K}}=1/3$ [panel (b)] (using $L=48$ sites). As described in the main text, $\langle \boldsymbol{\kappa}^2_{\ell}\cdot\boldsymbol{\kappa}^2_{m}\rangle$ becomes a smooth function of distance for $n_{\mathrm{K}}=1/2$ for most considered values of the interaction $U$, particularly in the intermediate region where this correlation is robust, indicative of a spiral that preserves the building blocks of the two-down two-up state. On the other hand, the $n_{\mathrm{K}}=1/3$ case exhibits three-site oscillations in a ``two-one'' pattern even when this correlation is robust in value. This is compatible with the three-down three-up block state of this density.

%--------------------------------------------------------------------------------
%--------------------------------------------------------------------------------

%================================================================================
\newpage
\begin{center}
{\bf \uppercase{Supplementary Note 2}\\Dimer correlation}
\end{center}

\begin{figure}[!htb]
\includegraphics[width=0.5\textwidth]{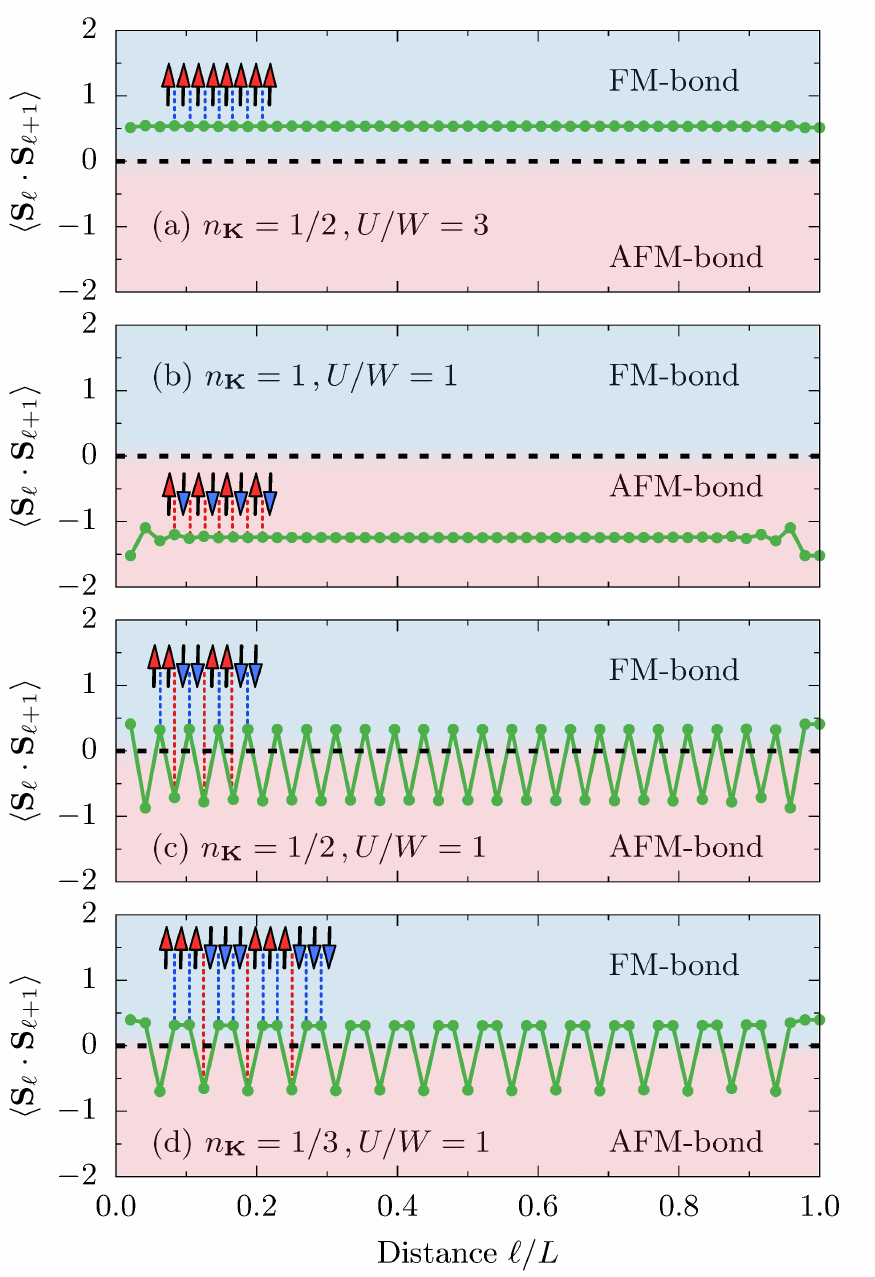}
\caption{Spatial dependence of the bond order operator \mbox{$\langle \mathbf{S}_{\ell}\cdot\mathbf{S}_{\ell+1}\rangle$} for various states: (a) FM state, (b) AFM state, (c) $\pi/2$-block state, and (c) $\pi/3$-block state. All results are calculated using the generalized Kondo-Heisenberg model on $L=48$ sites.}
\label{order}
\end{figure}

In Fig.~\ref{order} we present the bond order operator \mbox{$\langle\mathbf{S}_{\ell}\cdot\mathbf{S}_{\ell+1}\rangle$}. As described in the main text, this local dimer correlation $D_{\mathrm{order}}$ allow us to visualize the number of AFM and FM bonds in the states investigated. It is evident from the results depicted in Figs.~\ref{order}(c) and (d) that the block-magnetic order can be accurately identified by $D_{\mathrm{order}}$.

%--------------------------------------------------------------------------------
%--------------------------------------------------------------------------------

%================================================================================
\newpage
\begin{center}
{\bf \uppercase{Supplementary Note 3}\\Size dependence}
\end{center}

\begin{figure}[!hb]
\includegraphics[width=0.95\columnwidth]{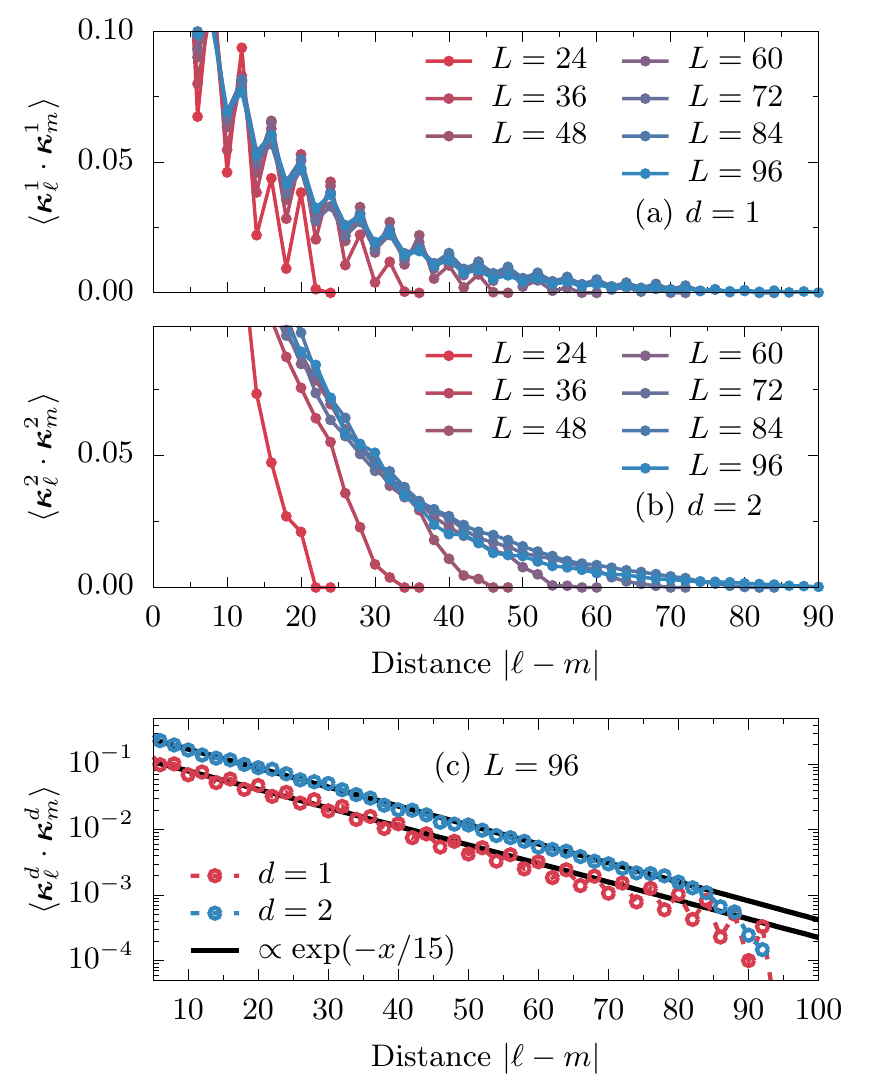}
\caption{System-size dependence of the chirality correlation function. (a) Nearest-- $d=1$ and (b) next--nearest $d=2$ neighbour chirality correlation $\langle \boldsymbol{\kappa}^d_\ell \cdot \boldsymbol{\kappa}^d_m\rangle$ calculated for various system sizes $L=24\,,\dots,96$. We use the generalized Kondo-Heisenberg model for $n_{\mathrm{K}}=1/2$ and $U/W=2.0$. (c) Log-plot illustrating the distance dependence of the chirality correlation $\langle \boldsymbol{\kappa}^d_\ell \cdot \boldsymbol{\kappa}^d_m\rangle$ function as calculated for $L=96$. The solid line represents a fit to the function $f(x)\propto \mathrm{exp}(-x/\xi)$.}
\label{size}
\end{figure}

In Figs.~\ref{size}(a,b), we present the system-size dependence of the nearest-- and next--nearest neighbour ($d=1$ and $d=2$, respectively) chirality correlation function $\langle\boldsymbol{\kappa}^{d}_{\ell}\cdot\boldsymbol{\kappa}^{d}_{m}\rangle$ for systems up to $L=96$ sites. Our results [see Fig.~\ref{size}(c)] suggest that the $\boldsymbol{\kappa}$-correlators decay exponentially with a robust correlation length $\xi\simeq 15$, where
\begin{equation}
\langle\boldsymbol{\kappa}^{d}_{\ell}\cdot\boldsymbol{\kappa}^{d}_{\ell+x}\rangle\propto\exp(-x/\xi)\,.
\end{equation}

\begin{figure}[!htb]
\includegraphics[width=0.5\textwidth]{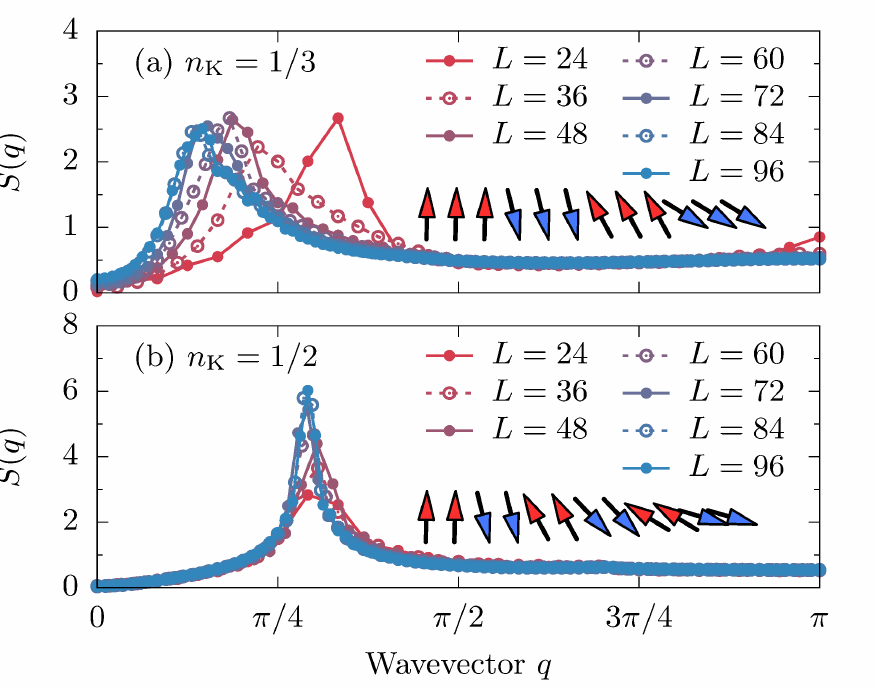}
\caption{System-size dependence of the spin structure factor $S(q)$. Panel (a) presents data for $n_{\mathrm{K}}=1/3$ and $U/W=1.2$ and (b) for $n_{\mathrm{K}}=1/2$ and $U/W=2$. Shown are results for various system sizes $L=24,\dots,96$ using the generalized Kondo-Heisenberg model.}
\label{sqsize}
\end{figure}

In Fig.~\ref{sqsize} we present the system-size dependence of the static spin structure factor $S(q)=\langle \mathbf{S}_{-q}\cdot\mathbf{S}_{q}\rangle$. In agreement with results for the chirality correlation functions, we find indications of short-range order behaviour because the peak strength does not continue growing with increasing $L$, but instead converges to a fixed value. Note that spirals with longer wavelengths (as in the case presented in Fig.~\ref{sqsize}(a) for $n_{\mathrm{K}}=1/3$) have stronger finite-size effects: this is natural because here the elementary building blocks involve three spins, instead of only two at $n_{\mathrm{K}}=1/2$. 

%--------------------------------------------------------------------------------
%--------------------------------------------------------------------------------

%--------------------------------------------------------------------------------
%--------------------------------------------------------------------------------

%================================================================================
%\clearpage
\begin{center}
{\bf \uppercase{Supplementary Note 4}\\Multi-orbital results}
\end{center}

\begin{figure}[!htb]
\includegraphics[width=0.5\textwidth]{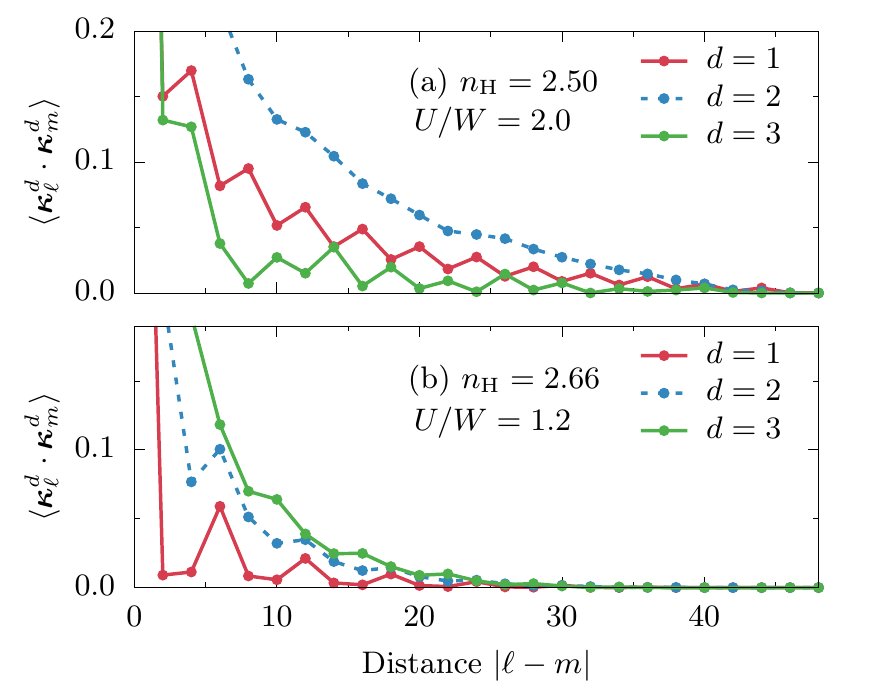}
\caption{Chiral correlation functions $\langle \boldsymbol{\kappa}^d_\ell \cdot \boldsymbol{\kappa}^d_m\rangle$
with $\boldsymbol{\kappa}^d_\ell=\mathbf{S}_\ell\times\mathbf{S}_{\ell+d}$. (a) Dependence of these chiral correlation functions vs distance between spins. Results calculated for the two--orbital Hubbard model, $L=48$, $n_{\mathrm{H}}=2.50$, and $U/W=2.0$. Panel (b): same as (a) but for $n_{\mathrm{H}}=2.66$ and $U/W=1.2$.}
\label{twoorb}
\end{figure}

In Fig.~\ref{twoorb} we repeat the main findings of our work, namely the nearest-- and next--nearest neighbour chirality correlation functions as shown for example in Fig.~4 of the main text, but now using the more complete, and more difficult, two--orbital Hubbard model,
\begin{eqnarray}
H_{\mathrm{H}}&=&-\sum_{\gamma,\gamma^\prime,\ell,\sigma}
t_{\gamma\gamma^\prime}
\left(c^{\dagger}_{\gamma,\ell,\sigma}c^{\phantom{\dagger}}_{\gamma^\prime,\ell+1,\sigma}+\mathrm{H.c.}\right)+
\Delta\sum_{\ell}n_{1,\ell}\nonumber\\
&+&U\sum_{\gamma,\ell}n_{\gamma,\ell,\uparrow}n_{\gamma,\ell,\downarrow}
+\left(U-5J_{\mathrm{H}}/2\right)\sum_{\ell}n_{0,\ell}n_{1,\ell}\nonumber\\
&-&2J_{\mathrm{H}}\sum_{\ell}\mathbf{S}_{0,\ell} \cdot \mathbf{S}_{1,\ell}
+J_{\mathrm{H}}\sum_{\ell}\left(P^{\dagger}_{0,\ell}P^{\phantom{\dagger}}_{1,\ell}
+\mathrm{H.c.}\right)\,.
\label{subhamhub}
\end{eqnarray}
In agreement with the data presented in Fig.~4 of the main text, the present results obtained with the multi-orbital system are fully consistent with the discussed block-spiral state displayed in the main text.

%--------------------------------------------------------------------------------
%--------------------------------------------------------------------------------

Let us now comment on other multi-band models. In particular, we will consider the following Hamiltonians closely related to the generalized Kondo-Heisenberg model:
\begin{itemize}
\item Kondo lattice (KL) --- $U=K=0$ in gKH,
\item Kondo Heisenberg (KH) --- $U=0$ in gKH.
\end{itemize}
Furthermore, we will investigate the gKH model where only one system parameter (either the Hund exchange $J_{\mathrm{H}}$ or the interaction $U$) is changed. Finally, we will present results for the full three-orbital Hubbard model \cite{Rincon2014,Herbrych2018,Patel2019} which exhibits OSMP with one localized and two itinerant bands. In the latter, we use $t_{00}=t_{11}=-0.5\,,t_{22}=-0.15\,,t_{01}=t_{10}=0\,,t_{02}=t_{20}=0.1$ together with $\Delta_0=-0.1\,,\Delta_1=0\,,\Delta_2=0.8$, all in units of eV. For all considered systems we fix the units to the kinetic energy bandwidth of gKH, i.e., $W=2.1\,\mathrm{eV}$. Some additional comments are necessary: (i) it is important to note that our KL and KH models have a factor of $2$ in front of the Hund exchange -- remnant of the original derivation of gKH. (ii) For the KH model, we fixed the value of the spin exchange to $\mbox{$K=4t_{11}^2/(2W)=0.0214$}$. (iii) For the results of gKH 
with only one variable being changed at a time we fix the remaining parameters to the value corresponding to $U/W=2$. (iv) For all Kondo-like models (KL, KH, gKH) we choose the $n_{\mathrm{K}}=1/2$ filling, namely $\pi/2$-block magnetic states in block-phase. The latter is stabilized \cite{Rincon2014,Herbrych2018} for $n_{\mathrm{H}}=4/3$ in the three-orbital Hubbard model which we use here.

\begin{figure}[!htb]
\includegraphics[width=0.5\textwidth]{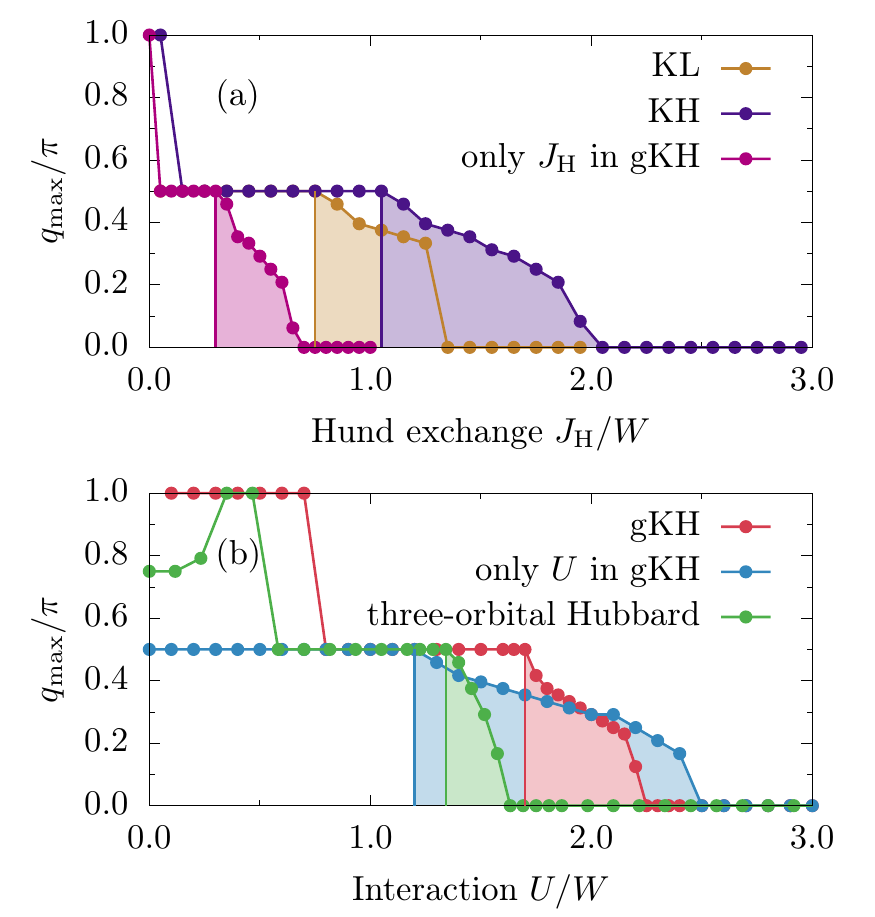}
\caption{Dependence of the maximum $q_{\mathrm{max}}$ of the static structure factor $S(q)$ on the value of (a) the Hund exchange $J_{\mathrm{H}}/W$ and (b) the interaction $U/W$. In (a) we present results for the Kondo lattice (KL, $U=K=0$), Kondo-Heisenberg (KH, $U=0\,,K=0.0214$), and generalized Kondo-Heisenberg (gKH) models with parameters fixed (to the value given by $U/W=2$) except the Hund exchange coupling. In (b) we present results for the full gKH, gKH with parameters fixed (to the value given by $U/W=2$) except interaction, and three-orbital Hubbard model (see text for details). Colored areas represent regions where spiral states are found.}
\label{phase_add}
\end{figure}

A good indicator of the spiral state in our investigation is the incommensurability of the maximum $q_{\mathrm{max}}$ of the static structure factor $S(q)$. In Fig.~\ref{phase_add} we present the Hund exchange and interaction $U$ dependence of $q_{\mathrm{max}}$ for various models. For KL we reproduce results~\cite{Garcia2004} where the spiral state is stabilized for $6\lesssim J_\mathrm{H}/t\lesssim 10$. Interestingly, introducing a finite exchange between localized spins $K\ne 0$ pushes the spirals to a larger range value of Hund exchange. Such a behaviour is an illustration of competing energy scales in our system: although the relatively small $K$ naively appears irrelevant, i.e. $K/J_{\mathrm{H}}\sim 0.01$, it has considerable effect on $S(q)$. The effect of finite interaction $U/W$ also modifies the dependence of spirals on Hund exchange. In the latter the relevant energy scale in reduced by a factor of $\sim4$.

Let us now comment on the effect of interaction $U/W$. As already shown in the main text, in the gKH model the block-spiral state can be found in the range $1.7\lesssim U/W\lesssim 2.3$. The two--orbital Hubbard model yields the same results \cite{Herbrych2019}. In the iron-based materials, for which our models are relevant, interaction $U/W$ can be naively controlled by, e.g., pressure. Assuming that other system parameters are not affected by such an action, the spiral region can be enlarged. In Fig.~\ref{phase_add}(b) we show that changes only in the interaction $U/W$ stabilizes block-spiral state in the range $1.2\lesssim U/W\lesssim 2.5$. Finally, to test the robustness of our findings we investigated the three-orbital Hubbard model. As is evident from Fig.~\ref{phase_add}(b), between the block and FM phases there is a region of incommensurate value of $S(q)$ and finite dimerization $D_{\pi/2}$ (not presented). As a consequence, the block-spiral state is evidently robust against modifications of the system parameters and even the number or orbitals, and thus we argue that it is a robust property of the OSMP itself.

%--------------------------------------------------------------------------------
%--------------------------------------------------------------------------------

%================================================================================
\clearpage
\begin{center}
{\bf \uppercase{Supplementary Note 5}\\Single-particle spectral function}
\end{center}

The single particle spectral function $A(q,\omega)=A^{\mathrm{e}}(q,\omega)+A^{\mathrm{h}}(q,\omega)$ is defined as
\begin{eqnarray*}
A^{\mathrm{e}}(q,\omega)&=&-\frac{1}{L} \sum_{\ell} \mathrm{e}^{i\ell q} \,\mathrm{Im}\langle \mathrm{gs}|c^{\dagger}_{\gamma,\ell}\frac{1}{\omega^{+}-H+\epsilon_0}c^{\phantom{\dagger}}_{\gamma,L/2}|\mathrm{gs}\rangle\,,\\
A^{\mathrm{h}}(q,\omega)&=&-\frac{1}{L} \sum_{\ell} \mathrm{e}^{i\ell q} \,\mathrm{Im}\langle \mathrm{gs}|c^{\phantom{\dagger}}_{\gamma,\ell}\frac{1}{\omega^{+}+H+\epsilon_0}c^{\dagger}_{\gamma,L/2}|\mathrm{gs}\rangle\,,
\end{eqnarray*}
where $c_{\gamma,\ell}=c_{\gamma,\ell,\uparrow}+c_{\gamma,\ell,\downarrow}$, $\omega^{+}=\omega+i\eta$, and $\eta=2\Delta\omega$ with $\Delta\omega=0.02$ as the frequency resolution. $A^{\mathrm{e}}(q,\omega)$ [$A^{\mathrm{h}}(q,\omega)$] represent retarded (electron photoemission) and advanced (hole inverse--photoemission) Green functions, respectively. The density of states can be calculated as $\mathrm{DOS}(\omega)=1/(\pi L)\sum_q A(q,\omega)$ and similarly for the electron and hole parts.

%================================================================================
\begin{center}
{\bf \uppercase{Supplementary Note 6}\\Frustrated long-range Heisenberg model}
\end{center}

In the main text, we have shown that qualitatively the block-chirality can be described effectively by the \mbox{$J_1$--$J_2$--$J_3$} Heisenberg model. It may be suspected that due to the oscillating nature of the mobile-electrons mediated spin-exchange, together with the long-range nature of spin correlations within OSMP, the generic effective spin-$1/2$ Hamiltonian that must be used to describe the discussed phenomena should be more extended and of the form
\begin{eqnarray}
\label{shamheis}
H&=&\sum_{\ell,r}\,J_r\,\mathbf{S}_\ell\cdot \mathbf{S}_{\ell+r}\,,\\
\text{with}\quad J_1<0\,,&\quad& J_2>0\,,\quad\mathrm{and}\quad J_{r>2}=|J_1|\frac{(-1)^{r-1}}{r^\alpha}\,.\nonumber
\end{eqnarray}
Such a frustrated long-range Hamiltonian is not suitable for DMRG-like approaches due to the well-known area law of entanglement which ``controls'' the accuracy of simulation. Thus, in order to gain understanding of the behaviour of the static structure factor $S(q)$ on such a model with extended interactions, we have used exact Lanczos diagonalization on a chain of $L=20$ sites with OBC. 

\begin{figure}[!htb]
\includegraphics[width=0.5\textwidth]{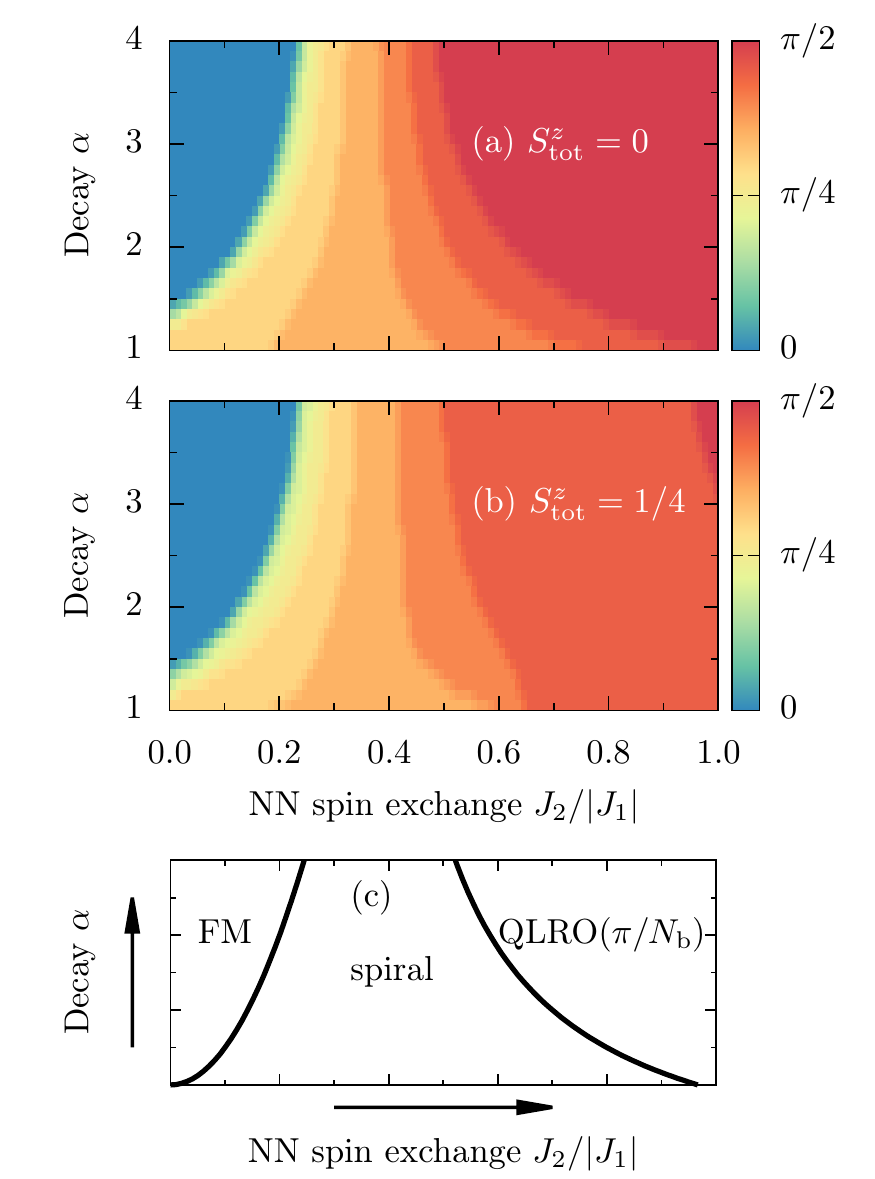}
\caption{Position of the maximum in the spin structure factor, $q_{\mathrm{max}}$, for the frustrated long-range Heisenberg model described in this Note. In panel (a) we preset results from magnetization $S^z_{\mathrm{tot}}=0$, while in (b) for $S^z_{\mathrm{tot}}=1/4$. Results were calculated using $L=20$ sites and the Lanczos diagonalization. (c) Schematic phase diagram of the frustrated long-range Heisenberg model.}
\label{longrange}
\end{figure}

In Fig.~\ref{longrange} we present the dependence of the position of the peak $q_{\mathrm{max}}$ on the $J_2$--interaction and on $\alpha$ in the Eq.~\eqref{shamheis} model. Panel (a) depicts results for overall magnetization $S^z_{\mathrm{tot}}=0$ while panel (b) for $S^z_{\mathrm{tot}}=1/4$. Depending on the magnetization, at large $J_2$ the system is dimerized with quasi-long range order (QLRO) of $q_{\mathrm{max}}=\pi/N_{\mathrm{b}}$ nature \cite{Onishi2015}, where $N_{\mathrm{b}}$ is the size of the block (e.g. $N_{\mathrm{b}}=2$ for $S^z_{\mathrm{tot}}=0$). On the other hand, in the $J_2\to 0$ limit the usual FM-order is stabilized. In between those two phases, similarly as in the case of the gKH model, the maximum of the spin structure factor acquires incommensurate values. It is known \cite{Ueda2014} that within this region a spiral can develop. A generic phase diagram of such a scenario is presented in Fig.~\ref{longrange}(c).

In summary, in this Note 5 we show that the successive FM-spiral-block transitions of the effective $J_1$-$J_2$-$J_3$ model described in the main text do not occur only in a narrow region of parameters. Here we have shown that a more generic model, with only two parameters $J_2/|J_1|$ and $\alpha$, comfortably supports such a physics. For this reason, our conclusions do not require fine tuning of couplings but are broadly universal and should appear in many models, and hopefully also in associated real materials.

%--------------------------------------------------------------------------------
%--------------------------------------------------------------------------------

\clearpage
%================================================================================
\end{document}